\documentclass[12pt]{article}
\usepackage{amsfonts,amsthm,amsmath,amssymb,upgreek}
\numberwithin{equation}{section}
\newtheorem{proposition}{Proposition}
\usepackage[paper=letterpaper,margin=1in]{geometry}
\usepackage[export]{adjustbox}
\usepackage[numbers,sort&compress]{natbib}
\usepackage{color}
\usepackage[normalem]{ulem}
\usepackage{cancel}
\usepackage{graphicx}% Include figure files
\usepackage{dcolumn}% Align table columns on decimal point
\usepackage{bm}% bold math
\usepackage{hyperref}% add hypertext capabilities
\usepackage{float}
\usepackage{makecell}
\usepackage[caption=false]{subfig}
\hypersetup{
    hypertexnames=false,
    colorlinks=true,
    linktoc=page,    %set to all if you want both sections and subsections linked
    linkcolor=blue,
    urlcolor=magenta
}
\usepackage{tikz,mathrsfs}
\usetikzlibrary{decorations.pathmorphing, decorations.markings, arrows.meta}
\usetikzlibrary{shadings}
\newcommand{\be}{\begin{equation}}
\newcommand{\ee}{\end{equation}}
\newcommand{\bea}{\begin{eqnarray}}
\newcommand{\eea}{\end{eqnarray}}
\newcommand{\dd}{\text{d}}

\newcommand{\mM}{\mathcal{M}}
\newcommand{\mF}{\mathcal{F}}
\newcommand{\mD}{\mathcal{D}}
\newcommand{\mL}{\mathcal{L}}
\newcommand{\mO}{\mathcal{O}}
\newcommand{\mS}{\mathbb{S}}
\newcommand{\bg}{\bar{g}}
\newcommand{\bn}{\bar{\nabla}}
\newcommand{\eff}{\text{eff}}
\newcommand{\rg}{r^{+}}
\newcommand{\rl}{r^{-}}

\newcommand{\daggerfootnote}[1]{%
    \renewcommand{\thefootnote}{\fnsymbol{footnote}}%
    \footnote[2]{#1}
    \renewcommand{\thefootnote}{\arabic{footnote}}%
}
\begin{document}
%===============================

\thispagestyle{empty}

\vspace*{.5cm}
\begin{center}

{\bf {\Large Universal Lichnerowicz Lifting \\
of Near-Horizon Soft Modes
%On the universality of quantum correction of black hole in the extremal limits
}\\
\vspace{1cm}}

 {\bf Peng Cheng\daggerfootnote{p.cheng.nl@outlook.com} and Yu-Qi Liu} \\
  \bigskip \rm

\bigskip

Center for Joint Quantum Studies and Department of Physics, \\School of Science, Tianjin University, Tianjin 300350, China\\

\rm

\vspace{2.5cm}
{\bf Abstract}
\end{center}

\begin{quotation}
A recurring universality appears in the low-temperature quantum thermodynamics of near-extremal black holes, where distinct parent geometries often lead to the same logarithmic temperature dependence at one loop.
In this work, we study the Lichnerowicz spectral origin of this infrared universality and understand why the relevant spectral data become insensitive to the details of the parent geometry.
For extremal near-horizon geometries containing a two-dimensional maximally symmetric throat, we construct the normalizable transverse-traceless tensor zero modes associated with near-horizon reparametrizations.
Turning on a small temperature lifts these zero modes through the first-order deformation of the Lichnerowicz operator.
Although the local matrix element depends on detailed parent-geometry data, these data cancel after projection onto normalized tensor modes, leaving the universal result.
For static spherically symmetric backgrounds, the resulting eigenvalue shift is proportional to the Fourier mode number and temperature.
For rotating backgrounds, the same structure persists in the near-horizon reparametrization tensor sector, while angular warp factors only modify the overall projection factor.
We fix the normalized coefficient explicitly, showing how radial integration and bulk normalization remove the near-horizon Taylor coefficients.
We further show that this lifted bulk spectrum is the Lichnerowicz realization of the Schwarzian soft sector.
Thus, the universal one-loop temperature dependence in the tensor sector is traced to an infrared bulk-boundary matching between near-horizon tensor zero modes and boundary reparametrization dynamics.

\end{quotation}

\vspace{1cm}

\setcounter{page}{0}
\setcounter{tocdepth}{2}
\setcounter{footnote}{0}

%%%%%%%%%%%%%%%%%%%%%%%%%%%%%%%%%%%%%%%%%%%%%%%%%%%%%%%%%%%%%%%%%%%%%%%%%%%%%%%%%%%%%%%%%%%%%%%%%%%%
% MAIN BODY
%%%%%%%%%%%%%%%%%%%%%%%%%%%%%%%%%%%%%%%%%%%%%%%%%%%%%%%%%%%%%%%%%%%%%%%%%%%%%%%%%%%%%%%%%%%%%%%%%%%%

\newpage
\noindent\rule[-10pt]{\linewidth}{0.05em}\par
\tableofcontents
\noindent\rule[-10pt]{\linewidth}{0.05em}\par
%\pagebreak

%%%%%%%%%%%%%%%%%%%%%%%%%%%%%%%%%%%%%%%%%%%%%%%%%%%%%%%%%%%%%%%%%%%%%%%%%%%%%%%%%%%%%%%%%%%%%%%%%%%%

\section{Introduction}
\label{intro}

Black hole thermodynamics provides one of the sharpest windows into quantum gravity. The Bekenstein-Hawking entropy, Hawking radiation, and the Euclidean gravitational path integral suggest that black holes should be treated as genuine quantum statistical systems rather than merely classical solutions of Einstein's equations \cite{Bekenstein:1973ur,Bardeen:1973gs,Hawking:1975vcx,Gibbons:1976ue}. This viewpoint is especially powerful near extremality, where a black hole develops a long throat and the low-temperature thermodynamics becomes sensitive to infrared physics \cite{Banerjee:2010qc,Sen:2012cj,Sen:2012dw}.
The near-extremal regime therefore provides a natural setting in which universal quantum corrections can be traced directly to the dynamics of the near-horizon region.

Extremal black holes are special not only because they have a long near-horizon throat, but also because their low-temperature statistical description raises several sharp questions.
At fixed charge or angular momentum, an extremal black hole sits at the lower edge of the classical black hole spectrum and has vanishing Hawking temperature, while its horizon area can remain macroscopic.
The semiclassical entropy therefore appears to describe an enormous zero-temperature degeneracy, whose microscopic meaning is far from automatic in a non-supersymmetric theory.
Slightly above extremality, the usual thermal description is also delicate.
The energy carried by a typical Hawking quantum can be comparable to the excitation energy above extremality, and the low-energy spectrum may depend sensitively on whether there is a true mass gap, a dense continuum, or an exponentially fine set of levels \cite{Preskill:1991tb,Page:2000dk}.
Geometrically, the extremal limit produces a long near-horizon throat, often locally described by an AdS$_2$ factor.
The strict AdS region, however, is not an ordinary decoupled infrared sector, because finite-energy excitations are constrained in a subtle way and the throat can exhibit fragmentation or boundary-mode effects \cite{Maldacena:1998uz}.
Recent analyses based on the gravitational path integral and Jackiw-Teitelboim (JT) gravity show that quantum fluctuations of the near-horizon region become increasingly important at low temperature and can qualitatively modify the density of states and the interpretation of the extremal limit \cite{Iliesiu:2020qvm,Turiaci:2023wrh}.
Thus, extremal black holes are not merely convenient limits of classical solutions, they isolate precisely the infrared sector where the statistical meaning of black hole entropy and the validity of semiclassical thermodynamics are most sharply tested.

Recent developments have made this infrared picture more concrete in the gravitational path integral. For near-extremal charged black holes, the putative mass-gap scale is better understood as the symmetry-breaking scale of the approximate near-horizon $SL(2,\mathbb R)$ symmetry, and the low-temperature partition function is governed by JT gravity coupled to gauge fields rather than by a naive semiclassical spectrum with a hard gap \cite{Iliesiu:2020qvm}.
Related analyses have shown that extremal zero modes and their finite-temperature regularization play an essential role in one-loop logarithmic corrections to black hole thermodynamics \cite{Charles:2019tiu,Heydeman:2020hhw,Boruch:2022tno,Iliesiu:2022onk,Rakic:2023vhv}.
The thermodynamic structure of near-extremal black holes has also been studied from the low-temperature expansion of the first law and from the intrinsic mechanics of near-horizon extremal geometries \cite{Johnstone:2013firstlaw,Hajian:2013nheg}. More recently, one-loop determinants in near-extremal black holes have been approached from a Lorentzian quasinormal-mode representation, providing a complementary spectral perspective on the Euclidean heat-kernel computations \cite{Mukherjee:2024qnm}.
This zero-mode mechanism extends beyond the charged spherically symmetric case. In near-extremal Kerr black holes, the near-horizon extreme Kerr throat provides the rotating analogue of the near-AdS$_2$ region \cite{Bardeen:1999px}, and the normalizable zero modes that enter the logarithmic correction to extremal rotating entropy \cite{Sen:2012cj} are lifted by the leading finite-temperature deformation, producing the characteristic logarithmic temperature correction \cite{Kapec:2023ruw}. More general rotating black holes exhibit the same tensor-mode contribution, with the universality traced to the near-horizon $SL(2,\mathbb R)$ symmetry and to cancellations in the Lichnerowicz operator \cite{Maulik:2024dwq}.
Further examples show that the same structure is not tied to asymptotically flat or AdS boundary conditions. Charged de Sitter black holes have cold, Nariai, and ultracold extremal limits with AdS$_2$, dS$_2$, and flat two-dimensional near-horizon factors, respectively \cite{Maulik:2025mmt,Blacker:2025zca}. 
Recent work has established the same tensor-mode contribution in a broad class of near-extremal black holes \cite{PandoZayas:2026vbg}. 
Further examples, including accelerating, Einstein-Gauss-Bonnet, and higher-derivative three-dimensional black holes, point to the same near-horizon soft mode mechanism \cite{Xu:2025dku,Alvarado:2026kio,Acito:2026mmf}.

The examples above strongly suggest that the low-temperature lifting of the near-horizon zero modes is not an accident of a particular black hole solution, but part of a universal structure.
However, this universality is far from obvious by looking at the deformed Lichnerowicz operator.
The first-order shift of a Lichnerowicz eigenvalue should appear to depend on the detailed parent geometry, including the metric information, the finite-temperature deformation, the gluing and ultraviolet data, and in rotating cases, the angular warping of the near-horizon metric.
It is therefore not obvious why the lifted spectrum should forget these data after projection onto the relevant zero modes.
A general demonstration of the cancellation of these data is therefore needed.
In particular, it is important to understand why the lifted eigenvalues are insensitive to ultraviolet gluing data, and how the resulting bulk spectrum is connected to the Schwarzian reparametrization theory that governs the near-AdS$_2$ soft sector.
One should be able to see the connection directly at the level of the lifted Lichnerowicz spectrum. This is the question to be addressed in this paper.

In this paper, we demonstrate how ultraviolet gluing data cancel in the normalized Lichnerowicz matrix element for a broad class of stationary black holes and how this cancellation leads to the universal tensor contribution to the low-temperature partition function.
The main novelty of this work is not to identify another example of logarithmic low-temperature behavior, but to provide a direct bulk spectral mechanism for it.
%The main novelty of this work is to provide a direct bulk spectral mechanism for logarithmic low-temperature behavior.
We show that the finite-temperature lifting of near-horizon tensor zero modes is governed by a normalized Lichnerowicz matrix element in which the parent-geometry data cancel after projection. 
%Comparing with the examples, we determine which geometric data survive in the normalized lifting coefficient.
%For the static backgrounds, the near-horizon Taylor coefficients entering at this order cancel, leaving only the horizon scale.
We give a radial boundary representation of the projected matrix element and relate its quadratic kernel to the Schwarzian kernel through bulk mode normalization.
This gives a bulk realization of the Schwarzian soft sector and explains why the final eigenvalue shift is linear in $|n|$ and $T$, despite the much more complicated local form of the deformed operator.
We consider general static spherically symmetric geometries with metric function $f(r)$, together with stationary rotating cases in which an angular warp factor $\sigma(\theta)$ dresses the two-dimensional throat.
In the extremal limit, the relevant near-horizon geometries contain a two-dimensional maximally symmetric factor $\mathcal M_2$ and a transverse sphere, allowing also for the warped and fibered embeddings that arise in rotating black holes.
We focus on transverse-traceless tensor fluctuations supported on the two-dimensional throat.
These modes are generated by large diffeomorphisms of the near-horizon geometry and lie in the reparametrization coset $\mathrm{Diff}(S^1)/SL(2,\mathbb R)$.
The leading finite-temperature lifting is fixed by the infrared soft sector of the throat.
For static spherically symmetric black holes, the normalized tensor zero modes acquire the universal eigenvalue shift
\begin{equation}
  \delta\lambda_n = \frac{\varepsilon}{16G}\cdot\frac{|n|T}{r_i}, \qquad |n|\ge2 ,
\end{equation}
where $r_i$ is the degenerate horizon radius and $\varepsilon=+1,-1,0$ for $\mathcal M_2=\mathrm{AdS}_2,\mathrm{dS}_2,\mathrm{Mink}_2$, respectively.
For stationary rotating black holes, the same universal $|n|\cdot T$ dependence persists in the near-horizon reparametrization tensor sector; the angular warp factor only changes the overall projection factor.
We explicitly verify this angular projection mechanism across different backgrounds, and the results are recovered consistently.
Thus, the parent geometry fixes the macroscopic scale, the angular normalization, and the sign of the two-dimensional throat, but not the universal dependence on the Fourier mode number and the temperature.

The mechanism is that the projected Lichnerowicz matrix element reduces to an infrared boundary contribution from the asymptotic region of the two-dimensional throat.
The continuous gluing data of the parent black hole therefore drop out of the normalized eigenvalue shift.
In Schwarzian language, the same soft sector is written in the boundary reparametrization basis, where the quadratic kernel is proportional to $T n^2(n^2-1)$.
The bulk Lichnerowicz problem instead uses bulk-normalized tensor perturbations, and the normalization map from the boundary reparametrization amplitude to the bulk tensor mode converts the Schwarzian kernel into the linear $|n|$ lifting.
In this sense, the lifted tensor zero modes are the bulk Lichnerowicz spectral realization of the Schwarzian reparametrization mode.
This is the spectral version of the infrared universality seen in the partition function, and it is consistent with the Schwarzian description of near-AdS$_2$ gravity \cite{Maldacena:2016upp,Stanford:2017thb,Mertens:2022irh}.
The Schwarzian description makes clear that the infrared theory is controlled by an approximate reparametrization symmetry that is spontaneously broken by the throat geometry and explicitly broken by the departure from extremality.

The paper is organized as follows. 
In section \ref{sec2}, we review the near-horizon geometries and the Lichnerowicz operator for the class of backgrounds considered here. 
We further construct the tensor zero modes, turn on a small temperature deformation, and compute their first-order eigenvalue shifts in spherical and rotating examples, in section \ref{sec3}. 
In section \ref{sec4}, we formulate the universality statement and explain its symmetry origin in terms of the Schwarzian soft sector. 
We conclude in section \ref{con} with a discussion of possible extensions.

%%%%%%%%%%%%%%%%%%%%%%%%%%%%%%%%%%%%%%%%%%%%%%%%%%%%%%%%%%%%%%%%%%%%%%%%%%%%%%%%%%%%%%%%%%%%%%%%%%%%

\section{Metric fluctuation and the Lichnerowicz operator}
\label{sec2}

In this section, we consider the quadratic fluctuations of the metric around a static background of the form $\mathcal{M}_2 \times \mS^{d-2}$, where $\mathcal{M}_2$ can be AdS$_2$, dS$_2$, or Mink$_2$ depending on the extremal limit and $\mS^{d-2}$ is a $(d-2)$-dimensional sphere of constant radius. The rotating backgrounds below instead have warped and fibered near-horizon geometries.
Such product geometries arise naturally as the near-horizon limits of extremal black holes in a wide class of theories. 
We obtain the second-order variation of the action, which governs the linearized dynamics of gravitational fluctuations. The kernel of this quadratic form defines the Lichnerowicz operator $\Delta_L$, a second-order elliptic differential operator acting on the tensor fluctuations.
We present the explicit form of $\Delta_L$, and discuss its properties under the transverse-traceless gauge. This operator plays a central role in the one-loop path integral for the gravitational field, and its zero modes will be crucial for the analysis of quantum corrections to black hole thermodynamics in the subsequent sections.

\subsection{The background}
\label{bkg}

We consider a gravitational theory coupled to matter fields $\Phi$. The total action is taken to be the Einstein-Hilbert action, minimally coupled with the matter action,
\begin{equation}
\begin{split}
	S[g,\Phi] &=\frac{1}{16\pi G} \int d^dx \, \sqrt{-g} \, (R - 2\Lambda) + S_{\text{matter}}[\Phi, g_{\mu\nu}]\\
	&=\frac{1}{16\pi G} \int d^dx \, \sqrt{-g} \, \left[(R - 2\Lambda) + 16\pi G\cdot \mL_{\text{matter}}\right]\,,
\end{split}\label{action}
\end{equation}
where the cosmological constant $\Lambda$ can be positive, negative, or zero. 

\subsubsection{Spherical black holes}

First of all, let us consider static, spherically symmetric solutions in $d$ dimensions, whose metric can be written in the general form
\begin{equation}
ds^2 = -f(r) \, \dd t^2 + \frac{\dd r^2}{f(r)} + r^2 d\Omega^2_{d-2}\,.\label{s-metric}
\end{equation}
The precise form of the metric function $f(r)$ depends on the matter content. For instance, in four dimensions, with $\Lambda=0$ and Maxwell matter, the solution is the Reissner-Nordstr\"{o}m (RN) black hole, for which
\begin{equation}
	f(r)= 1 - \frac{2GM}{r} + \frac{q^2}{r^2}\,,
\end{equation}
where $M$ and $q$ denote the mass and electric charge, respectively.
In general, such a spacetime may possess multiple horizons, which we denote collectively by $r_i$. The Hawking temperature associated with a horizon $r_i$ is given by
\begin{equation}
	T_i = \frac{1}{4\pi} f'(r_i)\,,\label{TH}
\end{equation}
where the prime denotes differentiation with respect to the radial coordinate $r$. 

When the temperature of a given horizon vanishes, the near-horizon geometry takes a factorized form $\mM_2 \times \mS^{d-2}$. The two-dimensional factor $\mM_2$ is determined by the behavior of $f(r)$ near the horizon. To avoid coordinate singularities, it is convenient to adopt the ingoing Eddington-Finkelstein coordinate defined by
\begin{equation}
	\dd v = \dd t + f(r)^{-1} \dd r\,,
\end{equation}
in terms of which the metric becomes
\begin{equation}
	ds^2 = -f(r) \, \dd v^2 + 2\dd v \dd r + r^2 d\Omega^2_{d-2}\,.\label{metric1}
\end{equation}
In the vicinity of the extremal horizon located at $r = r_i$, we introduce a small parameter $\lambda$ and rescale the coordinates as \cite{Bardeen:1999px}
\begin{equation}
	v \to \frac{v}{\lambda}\,,\qquad r \to r_i + \lambda \rho\,.
\end{equation}
At the horizon, we have
\begin{equation}
	f(r_i)=0\,,\qquad f'(r_i)=0\,,
\end{equation}
so that the metric function expands as
\begin{equation}
	f(r)=\frac{1}{2}f''(r_i)\lambda^2\rho^2+\mathcal{O}(\lambda^3)\,.
\end{equation}

Taking the $\lambda \to 0$ limit, the metric \eqref{metric1} reduces to
\begin{equation}
	ds^2=\underbrace{\left(- \frac{1}{2} f''(r_i) {\rho}^2 \dd{v}^2 + 2 \dd{v} \dd{\rho} \right)}_{\mM_2} + \underbrace{r_i^2 d\Omega_{d-2}^2}_{\mS^{d-2}}\,. \label{metric2}
\end{equation}
The first parenthesis describes a two-dimensional spacetime $\mM_2$; one can verify that it is a Lorentzian manifold with constant Ricci scalar
\begin{equation}
	R_2 = -f''(r_i)\,,\label{ricciS}
\end{equation}
inherited from the higher-dimensional parent geometry \eqref{metric1}. The sign of $f''(r_i)$ therefore dictates the nature of $\mM_2$:
\begin{itemize}
  \item If $f''(r_i) > 0$, then $R_2 < 0$, and $\mM_2$ is an $\mathrm{AdS}_2$ geometry.
  \item If $f''(r_i) < 0$, then $R_2 > 0$, and $\mM_2$ is a $\mathrm{dS}_2$ geometry.
  \item If $f''(r_i) = 0$, the manifold $\mM_2$ becomes two-dimensional Minkowski space $\mathrm{Mink}_2$. In this case, for example the ultra-cold limit of the RN-dS$_4$ spacetime, the metric can be written as
  \begin{equation}
  	ds^2_{2}=2\dd v\dd \rho=-\dd T^2+\dd R^2\,, 
  \end{equation}
  via the coordinate transformation
  \begin{equation}
  	\sqrt{2}\,T = v - \rho,\qquad \sqrt{2}\,R = v + \rho.
  \end{equation}
\end{itemize}
The second term in \eqref{metric2} represents a $(d-2)$-dimensional sphere of constant radius $r_i$. Since the metric is a direct sum of the two factors, the full $d$-dimensional spacetime factorizes into a product of two independent geometries, $\mM_2 \times \mS^{d-2}$.

\subsubsection{Rotating black holes}
\label{sec:rotating-background}

Let us study the near-horizon geometry of a $4$-dimensional rotating black hole, demonstrating its warped product structure $\mathrm{AdS}_2 \times \mS^{2}$ \cite{Bardeen:1999px}. 
The metric for a $4$-dimensional rotating black hole in Boyer-Lindquist-type coordinates $(t, r, \theta, \phi)$ can be written as
\begin{equation}
	ds^2 = -\frac{\Delta_r}{\Sigma} \left( \dd t - \frac{a \sin^2\theta}{\Xi} \dd \phi \right)^2 + \frac{\Sigma}{\Delta_r} \dd r^2 + \frac{\Sigma}{\Delta_\theta} \dd \theta^2 + \frac{\Delta_\theta \sin^2\theta}{\Sigma} \left( a \dd t - \frac{r^2 + a^2}{\Xi} \dd \phi \right)^2,
\end{equation}
with 
\begin{equation}
	\Sigma = r^2 + a^2 \cos^2\theta.
\end{equation}
We can keep functions $\Delta_r$, $\Delta_\theta$ and $\Xi$ as general functions of coordinates $r$ and $a$. For the Kerr-dS$_4$ black hole, the functions can be denoted as
\begin{equation}
\Delta_r = (r^2 + a^2)\left(1 - \frac{r^2}{\ell^2}\right) - 2GMr,\quad \Delta_\theta = 1 + \frac{a^2}{\ell^2} \cos^2\theta,\quad \Xi = 1 + \frac{a^2}{\ell^2}. 	
\end{equation}
The cosmological constant is $\Lambda = 3/\ell^2$, with the dS radius $l$.

Extremality occurs when two horizons coincide, i.e., when the function $\Delta_r$ has a double root at $r = r_i$. For the Kerr-dS$_4$ black hole, the cold limit is when $r_-=r_+$, and the Nariai limit is when $r_+=r_c$.
The extremality condition requires
\begin{equation}
	\Delta_r(r_i) = 0, \quad \Delta_r'(r_i) = 0.
\end{equation}
Solving these equations simultaneously determines the extremal mass $M_{\text{ext}}$ and horizon radius $r_i$ in terms of parameters of the parent spacetime. 
While the explicit expressions can be complicated, the key point is that near $r = r_i$, we can expand $\Delta_r$ as
\begin{equation}
	\Delta_r(r) \approx \frac{1}{2} \Delta_r''(r_i) (r - r_i)^2 + \mathcal{O}((r - r_i)^3).
\end{equation}
The near-horizon limit is achieved by taking $\lambda \to 0$ while simultaneously rescaling the time coordinate to focus on the region near the horizon.
For convenience, we define new coordinates $(\hat{t},\rho)$ by introducing a small scaling parameter $\lambda$
\begin{equation}
t=\frac{2(r_i^2+a^2)}{\lambda\Delta_r''(r_i)}\,\hat t,\qquad r=r_i+\lambda\rho,\qquad \phi=\hat\phi+\Omega_i t,\quad \Omega_i=\frac{a\Xi}{r_i^2+a^2}.
\end{equation}
The azimuthal shift places the limit in the frame corotating with the degenerate horizon.
At leading order
\begin{equation}
\Delta_r \approx \frac{1}{2} \Delta_r''(r_i)\lambda^2\rho^2\,.
\end{equation}

After a straightforward calculation, the near-horizon metric simplifies to a warped product of two-dimensional $\mM_2$ and a two-sphere $\mS^2$. The result can be written in the following general form
\begin{equation}
ds_{\text{NH}}^2 = \frac{2\Sigma_i}{\Delta_r''(r_i)}\left( -\rho^2 \dd\hat{t}^2 + \frac{\dd\rho^2}{\rho^2} \right) + \frac{\Sigma_i}{\Delta_\theta}\dd \theta^2+ \gamma(\theta) \left( \dd\hat{\phi} + e \rho \dd\hat{t} \right)^2.
\end{equation}
with
\begin{equation}
    	\Sigma_i = r_i^2 + a^2 \cos^2\theta\,,\quad \gamma(\theta) = \frac{\Delta_\theta(r_i^2+a^2)^2\sin^2\theta}{(r_i^2+a^2\cos^2\theta)\Xi^2}\,,\quad e = \frac{4ar_i\Xi}{(r_i^2+a^2)\Delta_r''(r_i)}
    \end{equation}
The first term is the metric of $\mM_2$ in Poincar\'e coordinates, warped by a function $ {2\Sigma_i}/{\Delta_r''(r_i)}$ that depends on the angular coordinate $\theta$. The Ricci scalar of $\mM_2$ can be derived as
\begin{equation}
	R_2 = -\frac{\Delta_r''(r_i)}{\Sigma_i}=-\left(\frac{\Delta_r}{\Sigma}\right)''\Bigg{|}_{r=r_i}\,.
\end{equation}
The left parts describe a deformed two-sphere $\mS^2$.
The constant $e$ is related to the angular velocity of the black hole at the horizon.

The near-horizon geometry of a general rotating black hole, characterized by the metric functions $\Delta_r$, $\Delta_\theta$, and $\Xi$, can be written as a warped product involving a two-dimensional factor $\mM_2$ and a deformed two-sphere. Depending on the sign of $\Delta_r''(r_i)$, the two-dimensional geometry $\mM_2$ is locally AdS$_2$, dS$_2$, or Mink$_2$. 
Near the extremal horizon, the combination $\Delta_r/\Sigma$ separates into the radial function $\Delta_r$ and the angular warp factor $1/\sigma(\theta)$.
Thus, in the throat-supported tensor sector, the role played by the static metric function $f(r)$ is taken over by an effective radial factor dressed by $1/\sigma(\theta)$.
More precisely, one may write

\begin{equation}
\frac{\Delta_r}{\Sigma}\Bigg{|}_{r=r_i}=\frac{\Delta_r(r_i)}{r_i^2}\cdot \frac{1}{\sigma(\theta)}\equiv f(r_i)\cdot \frac{1}{\sigma(\theta)}\,.
\end{equation}
where the dimensionless angular factor $\sigma(\theta)$ encodes the warping of the two-dimensional throat. For the standard Kerr-type convention
\begin{equation}
\Sigma=r^2+a^2\cos^2\theta,
\end{equation}
the dimensionless factor can be written as
\begin{equation}
\sigma(\theta)
=
1+\frac{a^2}{r_i^2}\cos^2\theta .\label{sigma}
\end{equation}
Thus, up to this angular warp factor, $\Delta_r/\Sigma$ provides the effective radial function governing the local $\mM_2$ geometry.

In the following sections we focus on metric fluctuations supported on the two-dimensional factor, namely $h_{ab}$. Although the curvature scale of $\mM_2$ acquires an additional $\theta$-dependence through the warp factor, the local equation determining the two-dimensional tensor modes remains unchanged. The extra angular dependence does not modify the universal radial soft-mode structure; rather, it enters only through the subsequent angular projection and normalization. Therefore, once the dynamics on $\mM_2$ is understood, the effect of rotation can be incorporated as an angular dressing of the same universal near-horizon mechanism.

This section illustrates the general mechanism by which a (warped) product structure $\mathcal{M}_2 \times \mS^{d-2}$ emerges in the extremal limit, with $\mathcal{M}_2$ being either $\mathrm{AdS}_2$, $\mathrm{dS}_2$, or $\mathrm{Mink}_2$ depending on the sign of $f''(r_i)$. 
In the subsequent sections, we will only work with metric \eqref{metric2} containing function $f(r)$ derived from the spherical parent geometry to demonstrate the mechanism. 
The same analysis applies to the near-horizon reparametrization tensor sector of the rotating geometries considered below, with the angular dependence entering through the projection and normalization factors.
Actually, since the final results are covariant and are obtained by integrating over $\theta$ from 0 to $\pi$, the $\theta$-dependence becomes an overall constant. 

In the extremal limit, a broad set of near-horizon geometries contains a two-dimensional space (which may be AdS, dS, or flat) and a sphere, provided the appropriate extremal condition is imposed.
The known results contain Kerr-de Sitter black holes in the Nariai limit (where $\mathcal{M}_2 = \mathrm{dS}_2$) \cite{Mariani:2025hee} , accelerating black holes \cite{Xu:2025dku}, and many more examples that will be discussed later. 
This suggests that the subsequent analysis of metric perturbations and the Lichnerowicz operator applies broadly, regardless of the specific details of the parent black hole solution.

Note that the $(d-2)$-dimensional sphere (or the two-sphere in particular) does not significantly affect the core calculation. Therefore, in the following, we will focus on the $d=4$ case, where the product geometry is $\mathcal{M}_2\times \mathbb{S}^2$.

\subsection{The Lichnerowicz operator}

Let us consider the theory shown in \eqref{action}.
Denoting the background metric as $\bar{g}_{\mu \nu}$ and the perturbation as $h_{\mu \nu}$, the total metric can be written as
\begin{equation}
g_{\mu \nu} = \bar{g}_{\mu \nu} + h_{\mu \nu}. \label{per1}
\end{equation}
All index raising and lowering are performed with the background metric $\bar{g}_{\mu \nu}$ and its inverse $\bar{g}^{\mu \nu}$. We can also denote $h = \bar{g}^{\mu \nu}h_{\mu \nu}$ and $h^{\mu \nu} = \bar{g}^{\mu \rho}\bar{g}^{\nu \sigma}h_{\rho \sigma}$.

We will mainly focus on metric perturbation $h_{\mu\nu}$. Now we can consider the effective action for $h_{\mu \nu}$. Due to the equation of motion, the first-order perturbation of the action is always zero, and we have
\begin{equation}
	S[\bar{g}+h, \Phi]=S[\bar{g}, \Phi]+\frac{1}{2}\cdot\frac{1}{16\pi G}\int d^dx\sqrt{-\bar g}~h_{\mu\nu}\cdot \Delta_L^{\mu\nu\rho\sigma}\cdot h_{\rho\sigma}+\mO(h^3)\,.
\end{equation}
The standard Hessian factor $1/2$ is retained; together with the Einstein-Hilbert prefactor in \eqref{action}, it gives the quadratic-action coefficient $1/(32\pi G)$.
The main task of the current subsection is to derive the operator $\Delta_L^{\mu\nu\rho\sigma}$. We need to expand everything to the second order of perturbation $h_{\mu\nu}$.

With the metric perturbation \eqref{per1}, we can further expand the inverse metric. Using the identity $g^{\mu \rho}g_{\rho \nu} = \delta_{\nu}^{\mu}$ and solving order by order, one can get
\begin{equation}
g^{\mu \nu} = \bar{g}^{\mu \nu} - h^{\mu \nu} + h^{\mu \rho}h_{\rho}^{\nu} + \mathcal{O}(h^3).
\end{equation}
The expansion of the tensor density $\sqrt{-g}$ can be derived using the expansion formula for matrix determinants. We have 
\begin{equation}
\sqrt{-g} = \sqrt{-\bar{g}}\left[1 + \frac{1}{2}h + \frac{1}{8}h^2 - \frac{1}{4}h_{\mu\nu}h^{\mu\nu} + \mathcal{O}(h^3)\right].
\end{equation}

The perturbations of the Christoffel symbols and curvature tensors are relatively complicated but straightforward. We can use $\delta$ and $\delta^2$ to denote the first-order and the second-order perturbations. 
We can briefly sketch the logic here.
For the Christoffel symbols, we can use the expression of the covariant derivative and write it in terms of the covariant derivative
\begin{equation}
	\Gamma_{\mu\nu}^{\lambda} =\bar{\Gamma}_{\mu\nu}^{\lambda} +\delta \Gamma_{\mu\nu}^{\lambda} +\delta^2\Gamma_{\mu\nu}^{\lambda} 
\end{equation}
with the first-order and second-order perturbations
\begin{equation}
\begin{split}
	\delta \Gamma_{\mu\nu}^{\lambda} &= \frac{1}{2} \bar{g}^{\lambda\rho} \left( \bar{\nabla}_\mu h_{\nu\rho} + \bar{\nabla}_\nu h_{\mu\rho} - \bar{\nabla}_\rho h_{\mu\nu} \right).\\
	\delta^2 \Gamma_{\mu\nu}^{\lambda} &= -\frac{1}{2} h^{\lambda\rho} \left( \bar{\nabla}_\mu h_{\nu\rho} + \bar{\nabla}_\nu h_{\mu\rho} - \bar{\nabla}_\rho h_{\mu\nu} \right).
\end{split}\label{Chris}
\end{equation}

The perturbations of the Ricci tensor can also be separated into the first-order and second-order parts, which are denoted by $\delta R_{\mu\nu}$ and $\delta^2 R_{\mu\nu}$.
The first-order Ricci perturbation only gets contributions from $\delta \Gamma^{\lambda}_{\mu\nu}$, and we have
\begin{equation}
\delta R_{\mu\nu} = \bar{\nabla}_{\lambda} \delta \Gamma^{\lambda}_{\mu\nu} - \bar{\nabla}_{\nu} \delta \Gamma^{\lambda}_{\mu\lambda}.\label{Rmn0}
\end{equation}
The second-order Ricci perturbation gets contributions from the following terms
\begin{equation}
\delta^2 R_{\mu\nu} = \bar{\nabla}_{\lambda} \delta^2 \Gamma^{\lambda}_{\mu\nu} - \bar{\nabla}_{\nu} \delta^2 \Gamma^{\lambda}_{\mu\lambda} + \delta \Gamma^{\lambda}_{\lambda\rho} \delta \Gamma^{\rho}_{\mu\nu} - \delta \Gamma^{\lambda}_{\nu\rho} \delta \Gamma^{\rho}_{\mu\lambda}.\label{RicciT2}
\end{equation}
Using $R = g^{\mu\nu}R_{\mu\nu}$, one can also get the first-order and second-order perturbations of the Ricci scalar, i.e., $\delta R$ and $\delta^2 R$. The first-order perturbation can be written as
\begin{equation}
\delta R= -h^{\mu\nu} \bar{R}_{\mu\nu} + \bar{g}^{\mu\nu} \delta R_{\mu\nu}.
\end{equation}
and the second-order perturbation as
\begin{equation}
\delta^2 R= h^{\mu}_{\ \lambda} h^{\lambda\nu} \bar{R}_{\mu\nu} - h^{\mu\nu} \delta R_{\mu\nu} + \bar{g}^{\mu\nu} \delta^2 R_{\mu\nu}.
\end{equation}
One can use \eqref{Rmn0} and \eqref{RicciT2} to write the explicit expression of the Ricci scalar perturbation.
For the cosmological constant term, the perturbation comes from the metric density.
We have
\begin{equation}
	\begin{split}
		\delta(\sqrt{-g}\Lambda)=\sqrt{-\bar{g}}~\frac{h \Lambda}{2}\,,\qquad \delta^2(\sqrt{-g}\Lambda)=\sqrt{-\bar{g}}\Lambda(\frac{1}{8}h^2 - \frac{1}{4}h_{\mu\nu}h^{\mu\nu} )\,.
	\end{split}\label{lambda}
\end{equation}

For the matter action shown in \eqref{action},
we can rewrite the first-order perturbation as
\begin{equation}
	\delta S_{\text{matter}}=\frac{1}{2}\int d^4 x\sqrt{-\bar{g}}~h^{\mu\nu}\bar{T}_{\mu\nu}\,,
\end{equation}
according to the definition of the energy-momentum tensor.
The second-order perturbation can be derived as
\begin{equation}
	\delta^2 S_{\text{matter}} = \frac{1}{4}\int d^4 x \sqrt{-\bar{g}} \left[  \delta T^{\mu\nu} h_{\mu\nu} + \frac{1}{2} h \, \bar{T}^{\mu\nu} h_{\mu\nu} \right].
\end{equation}
Here $\delta^2S_{\mathrm{matter}}$ denotes the coefficient of the term quadratic in $h_{\mu\nu}$ in the expansion.
Note that when the cosmological constant is regarded as matter, the corresponding energy-momentum tensor is $T_{\mu\nu}^{(\Lambda)}=-\frac{\Lambda}{8\pi G}g_{\mu\nu}$. The corresponding second-order term is
\begin{equation}
\delta^{2}S_\Lambda=-\frac{\Lambda}{8\pi G}\int d^4x\sqrt{-\bar g}\left(\frac18h^2-\frac14h_{\mu\nu}h^{\mu\nu}\right),
\end{equation}
which agrees with the perturbation of the cosmological constant part shown in \eqref{lambda}.
%In the Hessian convention of the quadratic action above, these two second-order contributions read
%\begin{equation}
%\delta^2 S_{\mathrm{matter}}=\frac{1}{32\pi G}\int d^4x\sqrt{-\bar g}\,8\pi G\left(\delta T^{\mu\nu}h_{\mu\nu}+\frac12h\,\bar T^{\mu\nu}h_{\mu\nu}\right),
%\end{equation}
%\begin{equation}
%\delta^2 S_\Lambda=\frac{1}{32\pi G}\int d^4x\sqrt{-\bar g}\,h_{\alpha\beta}\left(\Lambda\bar g^{\alpha\mu}\bar g^{\beta\nu}-\frac{\Lambda}{2}\bar g^{\alpha\beta}\bar g^{\mu\nu}\right)h_{\mu\nu}.
%\end{equation}
%Thus the $8\pi G$ matter coefficient in \eqref{eq:LL}, and the $\Lambda$ and $-\Lambda/2$ coefficients there, carry the same Hessian normalization.

Now, we can evaluate the second-order expansion of the action.
The gravitational density is $\mathcal{L} = \sqrt{-g}[(R - 2\Lambda)+16\pi G\cdot \mL_{\text{matter}}]$. Substituting the expansions derived above and keeping terms up to $h^{2}$, we have
\begin{equation}
\begin{aligned}
\mathcal{L} = \sqrt{-\bar{g}} \Bigg\{&
\underbrace{(\bar{R}-2\Lambda)+16\pi G\cdot \bar{\mL}_{\text{matter}}}_{\text{0th order}} \\
&+\underbrace{\left[ \delta R + \frac{1}{2}h(\bar{R}-2\Lambda)
+8\pi G \cdot h^{\mu\nu}\bar{T}_{\mu\nu}\right]}_{\text{1st order}} \\
&+\underbrace{\left[\begin{aligned}
\delta^2 R + \frac{1}{2}h \delta R
&+ \left(-\frac{1}{4}h_{\mu\nu}h^{\mu\nu} + \frac{1}{8}h^{2}\right)(\bar{R}-2\Lambda)\\
&+4\pi G \left(\delta T^{\mu\nu} h_{\mu\nu}
+ \frac{1}{2} h \, \bar{T}^{\mu\nu} h_{\mu\nu}\right)
\end{aligned}\right]}_{\text{2nd order}} \Bigg\} + \mathcal{O}(h^{3}).
\end{aligned}\label{action-order}
\end{equation}
The first-order terms vanish when the background satisfies the equations of motion, which is precisely the condition that the background is a solution of Einstein's equation
\begin{equation}
\bar{R}_{\mu\nu} - \frac{1}{2}\bar{R}\bar{g}_{\mu\nu} + \Lambda\bar{g}_{\mu\nu} = 8\pi G\bar{T}_{\mu\nu}.
\end{equation}
The left $\bar{g}^{\mu\nu} \delta R_{\mu\nu}$ term can be written as total derivative due to \eqref{Rmn0}.
The second-order term yields a quadratic form. After inserting the expressions for $\delta R$ and $\delta^2 R$, it yields the Lichnerowicz operator \cite{Maulik:2024dwq,Maulik:2025mmt}
\begin{equation}
\begin{split}
	& h^*_{\alpha\beta}~\Delta_L^{\alpha\beta\mu\nu} ~h_{\mu\nu} \\
	&= h^*_{\alpha\beta}\Bigg(\frac12 \bar{g}^{\alpha\mu}\bar{g}^{\beta\nu}\bar{\square}- \frac14 \bar{g}^{\alpha\beta}\bar{g}^{\mu\nu}\bar{\square}+ \bar{R}^{\alpha\mu\beta\nu} + \bar{R}^{\alpha\mu}\bar{g}^{\beta\nu} \\
& ~~~~~~~~~~~- \bar{R}^{\alpha\beta}\bar{g}^{\mu\nu}- \frac12 \bar{R}\,\bar{g}^{\alpha\mu}\bar{g}^{\beta\nu}+ \frac14 \bar{R}\,\bar{g}^{\alpha\beta}\bar{g}^{\mu\nu} \\
& ~~~~~~~~~~~+\Lambda \bar{g}^{\alpha\mu}\bar{g}^{\beta\nu} - \frac{\Lambda}{2}\bar{g}^{\alpha\beta}\bar{g}^{\mu\nu}\Bigg) h_{\mu\nu} +8\pi G\cdot\left[\delta T^{\mu\nu} h_{\mu\nu} + \frac{1}{2} h^{*}_{\alpha\beta}(\bar{g}^{\alpha\beta} \, \bar{T}^{\mu\nu}) h_{\mu\nu}\right]\,.
\end{split}\label{eq:LL}
\end{equation}
%Equation \eqref{eq:LL} is understood modulo longitudinal derivative terms, which vanish upon restriction to the transverse-traceless (TT) sector. Before gauge fixing, the complete metric Hessian contains additional longitudinal structures. All subsequent formulae refer to its TT projection.
There are gauge redundancies in the metric fluctuations, and the physical degrees of freedom can be described by modes satisfying the transverse-traceless (TT) gauge
\begin{equation}
	\nabla^{\mu} h_{\mu \nu} = 0, \quad h = 0. 
\end{equation}
In this gauge, the terms in \eqref{eq:LL} proportional to the traces $\bar{g}^{\alpha\beta}h^*_{\alpha\beta}$ or $\bar{g}^{\mu\nu}h_{\mu\nu}$ vanish. So we have
\begin{equation}
\begin{split}
h^*_{\alpha\beta} \Delta_L^{\alpha\beta\mu\nu} h_{\mu\nu}
&= \frac12 h^*_{\mu\nu} (\bar{\square}- \bar{R}+2\Lambda)h^{\mu\nu} \\
&\quad + h^*_{\alpha\beta} (\bar{R}^{\alpha\mu\beta\nu} +\bar{R}^{\alpha\mu}\bar{g}^{\beta\nu})h_{\mu\nu} + 8\pi G \cdot \delta T^{\mu\nu} h_{\mu\nu}\,.
\end{split}
\end{equation}
We can rewrite the operator as
\begin{equation}
\begin{split}
	& h^*_{\alpha\beta}~\Delta_L^{\alpha\beta\mu\nu} ~h_{\mu\nu} \\
	&= h^*_{\alpha\beta}\Bigg[\frac12 \bar{g}^{\alpha\mu}\bar{g}^{\beta\nu}\bar{\square}
	+ \bar{R}^{\alpha\mu\beta\nu}  + \left(\bar{R}^{\alpha\mu} - \frac12 \bar{R}\,\bar{g}^{\alpha\mu}+\Lambda \bar{g}^{\alpha\mu} -8\pi G\cdot\bar{T}^{\alpha\mu} \right)\bar{g}^{\beta\nu} \Bigg] h_{\mu\nu}\nonumber\\
	&~~+8\pi G\cdot \left( \delta T^{\mu\nu}h_{\mu\nu}+h^*_{\alpha\beta}\bar{g}^{\beta\nu}\bar{T}^{\alpha\mu}h_{\mu\nu} \right)\,.
\end{split}
\end{equation}
Combining Einstein's equation, the above Lichnerowicz operator simplifies to 
\begin{equation}
\begin{aligned}
h^*_{\alpha\beta}\Delta_L^{\alpha\beta\mu\nu}h_{\mu\nu}
&= h^*_{\alpha\beta}\left(\frac12 \bar{g}^{\alpha\mu}\bar{g}^{\beta\nu}\bar{\square}
+ \bar{R}^{\alpha\mu\beta\nu}\right)h_{\mu \nu}\\
&\quad+8\pi G\left( \delta T^{\mu\nu}h_{\mu\nu}
+h^*_{\alpha\beta}\bar{g}^{\beta\nu}\bar{T}^{\alpha\mu}h_{\mu\nu} \right).
\end{aligned}\label{Loperator}
\end{equation}
This is the operator that governs the dynamics of transverse-traceless gravitational perturbations.
Note that the second term in \eqref{Loperator} describes the contribution from matter fields. For scalar field $\phi$, it is
\begin{equation}
	-8\pi G\cdot h^*_{\alpha\beta} \left( \bar{g}^{\beta\nu}\partial^{\alpha}\phi\partial^{\mu}\phi\right)h_{\mu\nu}\,.
\end{equation}
It was shown in \cite{PandoZayas:2026vbg} that the matter fields obeying the extended Kunduri-Lucietti-Reall (KLR) near-horizon form do not contribute to the eigenvalue shift when considering a small temperature perturbation.
In the following, we restrict to matter sectors for which the first-order temperature deformation does not mix with the throat-supported transverse-traceless tensor sector.
This includes the extended KLR near-horizon form considered in the literature.
More general matter couplings, nonminimal interactions, or higher-derivative terms may modify the lifting and are outside the scope of the present analysis.
So, we will mainly consider the first three terms in \eqref{Loperator} in the current paper.

\section{Quantum corrections to black hole thermodynamics}
\label{sec3}

The thermodynamic entropy can be derived from the thermal partition function \cite{kapusta_finitetemperature_2023,Gibbons:1976ue}. The partition function can be evaluated from the Euclidean path integral by integrating over all possible configurations
\begin{equation}
	Z = \int [\mD g][\mD \Phi]\, e^{-I[g,\Phi]},
\end{equation}
where $I[g,\Phi]$ is the Wick-rotated Euclidean version of the action \eqref{action}. We mainly focus on the tensor fluctuation of the metric in this paper and have denoted $\bar{g}$ as the classical extremal metric solution of the theory. So the saddle-point contribution can be evaluated directly as
\begin{equation}
Z_{\text{cl}} =\exp\left(-\bar{I}[\bar{g},\bar{\Phi}]\right),\label{cl}
\end{equation}
where the action $\bar{I}$ corresponds to the zeroth-order contribution in \eqref{action-order}. 
When the metric fluctuation $h_{\mu\nu}$ is considered\footnote{Note that one can also consider perturbation of matter fields $\Phi=\bar{\Phi}+\phi$, and the corresponding path integral can be written as
\begin{equation}
	Z_{\phi}=\int[\mD \phi]~\exp\left[-\frac12\int \phi^*\Delta_M\phi\right]\,,
\end{equation}
which can be regarded as quantum field theory (QFT) on a fixed curved background \cite{Hawking:1975vcx,tHooft:1984kcu}.}, we can write the path integral up to the quadratic order as
\begin{equation}
	Z\approx Z_{\text{cl}}\times \int[\mD h]\exp\left[-\frac{1}{32\pi G}\int d^dx\sqrt{\bar{g}}~h^*_{\alpha\beta}~\Delta_L^{\alpha\beta\mu\nu} ~h_{\mu\nu}\right]\,,\label{pi}
\end{equation}
where $\Delta_L$ is the Lichnerowicz operator evaluated in the previous section.%\footnote{We only consider the one-loop correction from the metric perturbation in this paper. Unlike the approaches in the literature \cite{Iliesiu:2022onk,Xu:2025dku,Maulik:2024dwq,Blacker:2025zca,Maulik:2025mmt,PandoZayas:2026vbg}, we treat the QFT on a curved spacetime as the leading-order contribution, as shown in \eqref{cl}. The second-order perturbation action shown in \eqref{pi} also contains matter $\Phi$, but this action is too small compared with $\bar{I}$. We can evaluate the matter path integral using $\bar{I}$, and treat the matter field in \eqref{pi} as a fixed configuration. }.

The integral in \eqref{pi} is a Gaussian integration, which should give the determinant of the Lichnerowicz operator. Note that the zero modes should be dealt with separately, because a zero eigenvalue would spoil the determinant. For the non-zero modes, the partition function can be written as
\begin{equation}
	Z_{\text{n.z.}}=(\det \mathcal{D}_L)^{-1/2}\,,
\end{equation}
with $\mathcal D_L\equiv\Delta_L/(32\pi G)$.
One can deal with the determinant of the Lichnerowicz operator using the standard method \cite{berline2004heat,Ackermann:1994pg,Iochum:1996ph}. 

\subsection{Zero modes}

The zero modes arise due to the diffeomorphism that annihilates the Lichnerowicz operator $\Delta_L$.
We are mainly interested in the tensor zero modes on $\mM_2$, which can be solved explicitly. Putting the tensor zero modes into the Euclidean path integral, one can derive the corresponding partition function of the zero modes.

The Euclidean geometry we are studying is $\mM_2\times\mS^{2}$, which means that we can decompose the metric fluctuation $h$ into 
\begin{equation}
	h_{\mu\nu}\dd x^\mu \dd x^\nu=h_{ab}\dd x^a \dd x^b+h_{ij}\dd x^i \dd x^j\,.
\end{equation}
The coordinates $x^a$ are on the two-dimensional manifold $\mathcal{M}_2$ and coordinates $x^i$ on $\mS^{2}$. 
The metric perturbations $h_{\mu\nu}$ satisfies the the transverse-traceless (TT) gauge
\begin{equation}
h^\mu_{\;\mu}=0,\qquad \nabla^\mu h_{\mu\nu}=0 .
\end{equation}
We are mainly interested in perturbations that only excite the $\mathcal{M}_2$ part of the metric, i.e.
\begin{equation}
h_{ij}=0,\qquad h_{ai}=0 .
\end{equation}
Thus the only non-zero components are $h_{ab}=h_{ab}(x^a)$, which are functions on $\mathcal{M}_2$ only. We now translate the TT gauge into equations for $h_{ab}$.
The traceless condition can be rewritten as
\begin{equation}
0 = \bar{g}^{\mu\nu}h_{\mu\nu}= \bar{g}^{ab}h_{ab} + \bar{g}^{ij}h_{ij}= \bar{g}^{ab}h_{ab}.
\end{equation}
So $h_{ab}$ is traceless with respect to the two-dimensional metric $\bar{g}_{ab}$.
The transverse condition $\bar{\nabla}^\mu h_{\mu\nu}=0$ can be separated into $\nu=a$ and $\nu=i$.
Recall that the full covariant derivative $\bar{\nabla}$ decomposes into the direct product of the covariant derivatives on each factor. For any tensor with mixed indices, one uses the Levi-Civita connection of the product metric. 
In particular, for the field that depends only on $\mathcal{M}_2$ coordinates, the transverse condition can be written as $\bn^\mu h_{\mu a}=0$. Since the only non-zero component of $h_{\mu a}$ is when $\mu = b$, we have
\begin{equation}
\bn^\mu h_{\mu a}= \bn^b h_{ba}= \bg^{bc}\bn_c h_{ba}.
\end{equation}
Setting this to zero gives
\begin{equation}
\bn^a h_{ab}=0 .
\end{equation}
So $h_{ab}$ is also divergence-free on $\mathcal{M}_2$.
The condition for $\nu=i$ is automatically satisfied.
So, under the assumption that only the $\mathcal{M}_2$ components of the metric perturbation are non-zero, the full transverse-traceless conditions are equivalent to
\begin{equation}
\bg^{ab}h_{ab}=0\quad\text{and}\quad \bn^a h_{ab}=0 .
\end{equation}
That is, $h_{ab}$ is a symmetric traceless transverse tensor field on $\mathcal{M}_2$. 

Now, we can focus on the tensor modes that are diffeomorphisms on $\mM_2$.
Tensor modes are constructed by considering a diffeomorphism generated by a vector field $\zeta$ on $\mM_2$. The corresponding metric perturbation is given by the Lie derivative of the background metric
\begin{equation}
h_{ab} = \mathcal{L}_\zeta \bg_{ab}, \qquad \zeta = \zeta^a \partial_a.
\end{equation}
The TT gauge conditions transfer to the following requirements.
The traceless condition is automatically fulfilled if we express $\zeta$ in terms of a scalar field $\Phi(x^a)$ via
\begin{equation}
\zeta^a = \epsilon^{ab} \bn_b \Phi,
\end{equation}
where $\epsilon_{ab}$ is the Levi-Civita tensor on $\mM_2$. Substituting this into the transverse condition leads to a differential equation for $\Phi$,
\begin{equation}
(\bn_a\bn^a + R_2) \Phi = 0,\label{trans}
\end{equation}
with $R_2$ the Ricci scalar of $\mathcal{M}$, shown in \eqref{ricciS}. 
Note that originally, the vector $\zeta^a$ generating the diffeomorphism has two components, and the traceless condition serves as a constraint that leaves $\Phi$ as the only remaining degree of freedom.
Then, the modes satisfying equation \eqref{trans} from the transverse condition encode the metric fluctuations. Solving this equation yields a basis of scalar functions $\Phi^{(n)}$, each of which generates a distinct metric perturbation $h_{ab}^{(n)}$.
There can be an infinite number of $\Phi^{(n)}$ modes, and we call them zero modes because they annihilate the operator $\Delta_L$.
Although these modes are generated by diffeomorphisms, they should not be divided out as ordinary gauge redundancies once the asymptotic boundary condition of the throat is fixed.
They correspond to reparametrizations of the near-horizon boundary.
The modes with $n=0,\pm1$ generate the exact $SL(2,\mathbb R)$ isometries and are therefore excluded.
The remaining modes parameterize the soft sector associated with the quotient $\mathrm{Diff}(S^1)/SL(2,\mathbb R)$.

%\newpage

The tensor modes are obtained by solving the equation \eqref{trans}, on the two-dimensional Euclidean manifold $\mM_2$. From \eqref{metric2}, the two-dimensional Lorentzian metric can be written as
\begin{equation}
	ds^2=- \frac{1}{2} f''(r_i) {\rho_i}^2 \dd{v}^2 + 2 \dd{v} \dd{\rho_i}\,.\label{metric00}
\end{equation}
With proper coordinate transformations, one can rewrite the Euclidean metric on $\mM_2$ as
\begin{equation}
	ds^2 =\frac{2}{f''(r_i)}\left[{\sinh^2\eta}~\dd\tau^2 + \dd\eta^2\right]\,,\label{metric0}
\end{equation}
where $\tau$ is the dimensionless Euclidean time, with period $2\pi$.
The Ricci scalar of the geometry can be evaluated as
\begin{equation}
	R_2=-f''(r_i)\,.
\end{equation}
For cases $f''(r_i)>0$, the metric describes $\mathbb H^2$. Euclidean dS$_2$ is $S^2$, which has no corresponding nontrivial global TT tensor modes. The dS expressions below are therefore understood as an algebraic continuation of the local AdS$_2$ matrix element.

The $f''(r_i)=0$ case is a separate degenerate limit.
Defining new coordinates with
\begin{equation}
	y^2 = {\cosh^2\eta},\qquad \tau = \tau    \label{coT}
\end{equation}
we have
\begin{equation}
\dd\eta^2=\frac{\dd y^2}{y^2-1} ,\qquad \sinh^2\eta \, \dd\tau^2=(y^2-1) \dd\tau^2.
\end{equation}
Substituting into metric \eqref{metric0} gives
\begin{equation}
ds^2 = \frac{2}{f''(r_i)}\left[(y^2-1)\dd\tau^2 + \frac{\dd y^2}{y^2-1}\right].\label{metric-2}
\end{equation}

Working with the above metric \eqref{metric-2}, the equation \eqref{trans} reduces to
\begin{equation}
\frac{1}{y^2-1}\partial_\tau^2\Phi + \partial_y\big((y^2-1)\partial_y\Phi\big) - 2\Phi = 0.
\end{equation}
Separating variables as $\Phi(\tau,y)=e^{in\tau}\varphi(y)$ yields the ordinary differential equation for $\varphi(y)$:
\begin{equation}
\frac{\dd}{\dd y}\left((y^2-1)\frac{\dd\varphi}{\dd y}\right) - \left(\frac{n^2}{y^2-1}+2\right)\varphi = 0.
\end{equation}
The regular solutions generating normalizable tensor modes (for $|n|\ge 2$) are
\begin{equation}
\varphi_n(y) \propto \left(\frac{y-1}{y+1}\right)^{|n|/2}(|n|+y).
\end{equation}
Thus, the scalar field $\Phi$ that generates tensor zero modes is
\begin{equation}
\Phi_n(\tau,y) = N_n\, e^{in\tau} \left(\frac{y-1}{y+1}\right)^{|n|/2}(|n|+y),\qquad |n|\ge 2,\label{Phi}
\end{equation}
where $N_n$ is a normalization constant. The modes $n=0,\pm1$ correspond to Killing vectors and are excluded.
These solutions provide the complete set of scalar fields that generate the tensor zero modes in the respective near-horizon geometries.

Now we compute the metric perturbation $h_{\mu\nu}$ generated by these scalar fields. The diffeomorphism vector is
\begin{equation}
\zeta^a_n = \epsilon^{ab}\bar{\nabla}_b\Phi_n,
\end{equation}
with the Levi-Civita tensor $\epsilon^{ab}$ on $\mathcal{M}_2$. For the metric \eqref{metric-2}, the metric perturbation is given by the Lie derivative
\begin{equation}
h_{ab} = \bar{\nabla}_a\zeta_b + \bar{\nabla}_b\zeta_a,
\end{equation}
and in $(y,\tau)$ coordinates, we can get
\begin{equation}
\begin{aligned}
h_{\tau\tau} &= 2i n\big[(y^2-1)\Phi' - y\Phi\big],\\[4pt]
h_{\tau y} &= \frac{n^2}{y^2-1}\Phi + (y^2-1)\Phi'',\\[4pt]
h_{yy} &=2i n\left[-\frac{\Phi'}{y^2-1} + \frac{y\Phi}{(y^2-1)^2}\right].
\end{aligned}
\end{equation}  
Substituting the expression of $\Phi$ shown in \eqref{Phi}, and using the differential equation of $\Phi$, we obtain
\begin{equation}
h_{\mu\nu}^{(n)} \dd x^\mu \dd x^\nu = -2 N_n |n|(n^2-1)
 e^{in\tau} \left(\frac{y-1}{y+1}\right)^{|n|/2} \left[ -\dd\tau^2 + \frac{2i\, \dd\tau \dd y}{y^2-1}  + \frac{\dd y^2}{(y^2-1)^2}  \right].\label{hmn0}
\end{equation}
Equation~\eqref{hmn0} are written for $n\geq2$; negative-frequency modes are defined by $h^{(-n)}=h^{(n)*}$.
$N_n$ can be determined by the orthonormality with respect to the inner product
\begin{equation}
	\langle h^{(n)}|h^{(m)}\rangle = \int d^4x\sqrt{\bar{g}}\; h_{\alpha\beta}^{(n)*}\bar{g}^{\alpha\mu}\bar{g}^{\beta\nu}h_{\mu\nu}^{(m)} = \delta^{nm}.\label{deltamn}
\end{equation}
The normalization constant $N_n$ can be expressed as
\begin{equation}
N_n = \frac{\sqrt{\varepsilon}}{4\pi r_i \sqrt{ |f''(r_i)|}\,\sqrt{|n|(n^2-1)}}
\label{Nn}
\end{equation}
where we have introduced $\varepsilon=\pm 1$ to highlight that $f''(r_i)$ can be negative for dS$\times \mS^2$ geometry.

Note that the metric perturbation is derived using the coordinate $y$, which is related to $\eta$ by $y=\cosh\eta$. It can then be transformed back to the coordinate system used in \eqref{metric0}.
For $\varepsilon>0$, which means that we work in the AdS case, the metric perturbation on metric \eqref{metric0} can be written as
\begin{equation}
	h_{\mu\nu}^{(n)}\dd x^\mu \dd x^\nu = \frac{ie^{in\tau}\ell_{\mathrm{AdS}}\sqrt{|n|(n^2-1)}}{\sqrt{8}\pi r_i}\tanh^{|n|}\frac{\eta}{2}\Bigl(-\dd \tau^2+\frac{2i\,\dd \tau\,\dd \eta}{\sinh\eta}+\frac{\dd \eta^2}{\sinh^2\eta}\Bigr),\label{hmn}
\end{equation}
which is transverse, traceless, and satisfies $\bar\nabla^a h^{(n)}_{ab}=0$ and the zero-mode equation at extremality. 
The result \eqref{hmn} is in agreement with the expression obtained in the literature \cite{Blacker:2025zca}.
The dS case can be regarded as the analytical continuation of the AdS case.

The Minkowski case is a bit tricky.
In the limit $\varepsilon\to0$, the differential equation is changed, and we obtain
	\begin{equation}
	\Phi_n\ \propto e^{in\tau}\,\eta^{|n|}.\label{flat}
	\end{equation}
Those modes are non-normalizable, and we therefore regard the corresponding tensor zero modes as absent from the regulated partition function. This is the strategy we will take in this paper.

\subsection{Turning on a small temperature}

From the previous part, we have obtained an infinite number of modes, while the effective action for those zero modes is zero. So the path integral is divergent, and one needs to regularize it by turning on a small temperature. 
This is what we will do in this subsection.

For a generic spherically symmetric black hole metric \eqref{s-metric} with metric function $f(r)$, the extremal horizons $r_i$ satisfy 
\begin{equation}
	f(r_i)=f'(r_i)=0\,.
\end{equation}
When we deform the extremal horizon by heating it a bit, the extremal horizon $r_i$ turns into two near-extremal horizons, denoted by $\rg_i$ and $\rl_i$.
Let us look at the outer one $\rg_i = r_i + \delta_i$ with small $\delta_i>0$.
The corresponding Hawking temperature $T_i$ is defined in \eqref{TH}, which is a small temperature.
Expanding $f'(\rg_i)$ around $r_i$ gives
\begin{equation}
f'(\rg_i) = f'(r_i) + f''(r_i)\delta_i + \mO(\delta^2) = f''(r_i)\delta_i + \mO(\delta^2).
\end{equation}
Hence, to leading order, the Hawking temperature can be expressed as
\begin{equation}
T_i \approx \frac{f''(r_i)}{4\pi}\,(\rg_i-r_i)\,.
\end{equation}
Defining
\begin{equation}
\alpha_i = \frac{4\pi}{f''(r_i)}\,,
\end{equation}
we can rewrite the outer horizon as
\begin{equation}
\rg_i=r_i+ \alpha_i\,T_i.\label{rg}
\end{equation}
The label $i$ is used to denote the horizons in different extremal limits, like for the RN-dS case, $r_i$ can be $r_{\text{cold}}$ in the cold limit and $r_{\text{N}}$ in the Nariai limit.
We will omit the label $i$ in the Hawking temperature for simplicity, as the temperature for the horizon under consideration is unambiguous.

Performing an analytic continuation $t = -i\tau/(2\pi T)$ and $\tau=2\pi Tt_E$, so that the Euclidean metric is
\begin{equation}
	ds_E^2 = \frac{f(r)}{(2\pi T)^2} \dd\tau^2 + \frac{\dd r^2}{f(r)} + r^2 d\Omega_{2}^2,
\end{equation}
To write the metric in the $\mM_2\times \mS^{2}$ form,  we can do the coordinate transformation
\begin{equation}
r = \rg_i + \alpha_i (y-1)T,\qquad y\in[1,\infty),\label{trans78}
\end{equation}
where $y=1$ corresponds to the outer horizon, one can expand the metric in powers of $T$.
Note that, as will be seen, the specific choice of $\alpha$ in \eqref{trans78} guarantees that when expanding $f(r)$ to order $T^2$, the leading terms in the metric take the simple hyperbolic form. We have
\begin{equation}
	\frac{\dd r}{\dd y}=\alpha_i \, T\,.
\end{equation}

Expanding $f(r)$ around $r=\rg_i$ and using \eqref{rg} yields
\begin{equation}
f(r) = A_2\cdot T^2 + A_3 \cdot T^3 + \mO(T^4),
\end{equation}
where $A_2$ and $A_3$ are
\begin{equation}
A_2 = \frac{8\pi^2}{f''(r_i)}(y^2-1),\qquad
A_3 = \frac{32\pi^3 f'''(r_i)}{3 f''(r_i)^3}(y-1)^2(y+2).
\end{equation}
To the subleading-order, the metric components are then expanded as
\begin{align}
g_{\tau\tau} &= \frac{f(r)}{(2\pi T)^2} = \frac{A_2}{(2\pi)^2} + \frac{A_3}{(2\pi)^2}\cdot T ,\label{83}\\
g_{yy} &= \frac{1}{f(r)}\left(\frac{\dd r}{\dd y}\right)^2 = \frac{\alpha_i^2}{A_2} - \frac{\alpha_i^2 A_3}{A_2^2}\cdot T ,\\[8pt]
g_{\theta\theta} &= r^2 = r_i^2 + 2r_i\alpha_i y \cdot T,\\[6pt]
g_{\phi\phi} &= r^2 \sin^2\theta = r_i^2 \sin^2\theta + 2r_i\alpha_i y  \sin^2\theta \cdot T.\label{86}
\end{align}
Substituting the expressions for $A_2$, $A_3$ and $\alpha_i$ gives the leading-order background metric
\begin{equation}
\bar{g}_{\mu\nu}\dd x^\mu \dd x^\nu = \frac{2}{f''(r_i)}\left((y^2-1)\dd\tau^2 + \frac{\dd y^2}{y^2-1}\right) + r_i^2 d\Omega_2^2,\label{barg}
\end{equation}
which is equivalent to the metric on $\mM_2$ in \eqref{metric2} as can be shown through proper coordinate transformation. 
And the first-order correction can be derived as
\begin{equation}
T\cdot \delta g_{\mu\nu}\dd x^\mu \dd x^\nu = T\left[\frac{8\pi f'''(r_i)}{3 f''(r_i)^3}(y+2)(y-1)^2 \dd\tau^2 - \frac{8\pi f'''(r_i)}{3 f''(r_i)^3}\frac{(y+2)\dd y^2}{(y+1)^2} + 2r_i\alpha_i y\, d\Omega_2^2\right].\label{deltag}
\end{equation}
For the special case of the RN-dS black hole, in the cold black hole limit, one has
\begin{equation}
f''(r_i)=\frac{2}{r_0^2}\left(1-\frac{6r_0^2}{\ell^2}\right),\qquad
f'''(r_i)=-\frac{12}{r_0^3}\left(1-\frac{4r_0^2}{\ell^2}\right),
\end{equation}
which reproduces the results in \cite{Maulik:2025mmt}.
\begin{equation}
	T\cdot \delta g_{ab}\dd x^a \dd x^b = T\cdot \frac{4\pi r_0^3 \ell^4\,(\ell^2-4r_0^2)}{(\ell^2-6r_0^2)^3}\left[-(y+2)(y-1)^2 \dd\tau^2+\frac{(y+2)\dd y^2}{(y+1)^2}\right].
\end{equation}

\subsection{Eigenvalue shift and the zero mode contribution}

From the previous subsection, by turning on a small temperature, the metric can be written as 
\begin{equation}
	g_{\mu\nu} = \bar{g}_{\mu\nu} + T \cdot\delta g_{\mu\nu},
\end{equation}
which are explicitly shown in equations \eqref{barg} and \eqref{deltag}.
Correspondingly, the Lichnerowicz operator also be corrected. To the first-order, it can be written as
\begin{equation}
	\Delta_L^{\alpha\beta\mu\nu}=\bar{\Delta}_L^{\alpha\beta\mu\nu} + \delta\Delta_L^{\alpha\beta\mu\nu}\,,
\end{equation}
where $\bar{\Delta}_L$ is the original Lichnerowicz operator shown in \eqref{Loperator}.
We include the action prefactor $1/(32\pi G)$ in $\delta\lambda_n$.
The Gaussian Hessian kernel is $\mathcal D_L=\Delta_L/(32\pi G)$, so $\delta\mathcal D_L=\delta\Delta_L/(32\pi G)$.
Applying the operator on $h_{\mu\nu}$, the eigenvalue can be expressed as $\lambda_n+\delta \lambda_n$, with the first-order correction to the eigenvalue
\begin{equation}
	\delta \lambda_n = \frac{1}{32\pi G}\int d^4x \sqrt{\bar{g}} \; h_{\alpha\beta}^{(n)*} \; \delta\Delta_L^{\alpha\beta\mu\nu} \; h_{\mu\nu}^{(n)}.\label{Ecorrection}
\end{equation}
For bosonic metric fluctuations, the Gaussian integral gives a determinant factor of the quadratic kernel; its temperature-independent normalization is absorbed into the measure,
\begin{equation}
	Z_{\rm n.z.}\propto (\det \mathcal D_L)^{-1/2}.
\end{equation}
The zero modes of the extremal operator must be treated separately.
After a small temperature is turned on, the lifted zero modes contribute
\begin{equation}
	\delta\log Z = -\frac12\sum_{|n|\geq2}\log \delta\lambda_n .
\end{equation}
Here and below, $\delta\log Z$ denotes only the TT reparametrization contribution. Rotational modes on $S^2$, $U(1)$ modes, matter, ghosts, and other boundary sectors must be included separately for a complete one-loop result.
The main task is to evaluate the eigenvalue shift shown in \eqref{Ecorrection}.

From \eqref{Loperator}, the gravitational part with the deformation $\delta g_{\mu\nu}$ is
\begin{equation}
	h^*_{\alpha\beta}~\Delta_L^{\alpha\beta\mu\nu} ~h_{\mu\nu} = h^*_{\alpha\beta}~\left(\frac12 {g}^{\alpha\mu}{g}^{\beta\nu}{\square}+ {R}^{\alpha\mu\beta\nu}
	 \right)~h_{\mu \nu}.
\end{equation}
Thus, with all inverse-metric variations included, the first-order deformation is
\begin{equation}
	h^*_{\alpha\beta}~\delta\Delta_L^{\alpha\beta\mu\nu} ~h_{\mu\nu} =  \frac12 h^*_{\alpha\beta}\,\delta\!\left(g^{\alpha\mu}g^{\beta\nu}\square\right)h_{\mu\nu} + h^*_{\alpha\beta}\delta R^{\alpha\mu\beta\nu} h_{\mu\nu} \,,\label{LO}
\end{equation}
The first term includes both $\delta\square$ and $\delta(g^{\alpha\mu}g^{\beta\nu})$, and no explicit cosmological-constant term remains in \eqref{Loperator}.
The result shown in \eqref{LO} is the same as the result obtained in \cite{Maulik:2024dwq,PandoZayas:2026vbg}.

\subsubsection*{Cancellation of parent-geometry data}

Now, we need to evaluate the eigenvalue shift
\begin{equation}
	\delta \lambda_n = \frac{1}{32\pi G}\int d^4x \sqrt{\bar{g}} \; \Big[ h^*_{\alpha\beta}~\delta\Delta_L^{\alpha\beta\mu\nu} ~h_{\mu\nu}\Big].\label{lambda-n}
\end{equation}
with $h^*_{\alpha\beta}~\delta\Delta_L^{\alpha\beta\mu\nu} ~h_{\mu\nu}$ shown in \eqref{LO}.
$h^*_{\alpha\beta}~\delta\Delta_L^{\alpha\beta\mu\nu} ~h_{\mu\nu}$ is generally depend on the metric perturbation $h^{(n)}_{\mu\nu}$ in \eqref{hmn}, and the deformed background $\bar{g}_{\mu\nu}+T\cdot \delta g_{\mu\nu}$ shown in \eqref{barg} and \eqref{deltag}.
So, one can expect $h^*_{\alpha\beta}~\delta\Delta_L^{\alpha\beta\mu\nu} ~h_{\mu\nu}$ to have a relative complicated dependence on $f''(r_i)$, $f'''(r_i)$ and $y$. For $n\geq2$, the result reads
\begin{eqnarray}\label{hDh}
	&& h^*_{\alpha\beta}~\delta\Delta_L^{\alpha\beta\mu\nu} ~h_{\mu\nu} =T \cdot\frac{8 \pi N_n^2 n^2(n^2-1)^2 f''(r_i) }{3 r_i}\frac{(y-1)^{n-2}}{(y+1)^{n+4}} \label{integrand}\\ \nonumber
	&&\times\Big[2 r_i f'''(r_i) \left(9 n-10 y+4-(y+2) (2 y^2+n y-2 n^2)\right)-3 (y+1)^2 (n-2 y) f''(r_i)\Big]\,.
\end{eqnarray}
Substituting the above result into \eqref{lambda-n} yields
\begin{equation}
	\delta \lambda_n=\frac{\varepsilon}{16 G}\frac{|n| T }{r_i}\,,\qquad |n|\ge 2 \label{106}
\end{equation}
where $\varepsilon=\pm 1$ comes from the normalization factor of the metric perturbation $N_n$, shown in \eqref{Nn}.
As discussed in \cite{Blacker:2025zca}, when the near-horizon geometry is EAdS$_2\times \mS^2$, one has $\varepsilon=1$; when the geometry contains EdS$_2$, one obtains $\varepsilon=-1$.
Because there are no normalizable tensor soft modes in the ultracold limit with a Mink$_2$ factor, the corresponding lifted zero-mode contribution is absent.
In such a sense, the $\varepsilon=0$ situation is also allowed, as a consistent limit of the AdS or dS cases.

For the EAdS case, we have $\varepsilon=1$.
Each mode contributes a Gaussian functional integral of the form $(\det\delta \lambda_n)^{-1/2}$. Summing over all modes with $|n|\ge 2$, the logarithmic correction to the partition function is therefore
\begin{equation}
\delta\log Z
= -\frac{1}{2}\sum_{|n|\ge 2}\log(\delta \lambda_n)
\end{equation}
Because the eigenvalues depend only on $|n|$, i.e. $\delta\lambda_{+n}=\delta\lambda_{-n}$, the sum over $|n|\ge 2$ receives equal contributions from $+n$ and $-n$. Hence
\begin{equation}
-\frac{1}{2}\sum_{|n|\ge 2}\log(\delta \lambda_n)
= -\sum_{n\ge 2}\log(\delta \lambda_n)
= -\log\prod_{n\ge 2}(\delta \lambda_n).
\end{equation}
Substituting \eqref{106} gives the formal infinite product
\begin{equation}
\prod_{n\ge 2}(\delta \lambda_n)
= \prod_{n\ge 2}\frac{n}{\alpha},
\qquad\text{with}\quad
\alpha\equiv\frac{16G\cdot r_{i}}{T}.
\end{equation}
The determinant is understood relative to a reference eigenvalue scale; changing this scale affects only the temperature-independent term.
The product $\prod_{n\ge 2}n$ is divergent and must be regularized. Using the zeta-function prescription, we can get
\begin{equation}
\prod_{n\ge 2}\frac{n}{\alpha}
= \alpha^{3/2}\sqrt{2\pi}.
\end{equation}
This result follows from the analytic continuation of the Riemann zeta function. %, $\zeta(0)=-1/2$, so that $\sum_{n\ge 2}1 = \zeta(0)-1 = -3/2$.
Inserting the regularized product back into the logarithm
\begin{align}
\delta\log Z
= \frac{3}{2}\log\!\left(\frac{T}{16G \cdot r_{i}}\right)-\frac{1}{2}\log(2\pi) = \frac{3}{2}\log\frac{T}{r_{i}}+\mO(\log G)\,.
\end{align}
Thus, the universal temperature-dependent part of the TT soft-sector contribution is
\begin{equation}
\delta\log Z
= \frac{3}{2}\log\frac{T}{r_{i}}+\cdots
\end{equation}
where the dots denote temperature-independent and higher-order terms. The result is that all model-dependent details of the near-horizon geometry cancel out in this lifted zero-mode sector, and the final answer depends only on the universal ratio $T/r_{i}$, reflecting the robustness of the throat dynamics.

For the case with $\varepsilon=-1$, for example the Nariai limit of the RN-dS$_4$ black hole, the  near-horizon geometry is $(-\mathrm{EAdS}_{2})\times \mS^{2}$.
The first-order eigenvalue correction of the graviton operator $\bar{D}$ becomes negative
\begin{equation}\label{eq:tensor-neg}
\delta\lambda_{n} = -\frac{|n|\,T}{16G \cdot r_{i}}  < 0\,.
\end{equation}
Consequently, the Gaussian factor $\exp\!\left[-\delta\lambda_{n}\langle h^{(n)}|h^{(n)}\rangle\right]$ grows exponentially, and the Euclidean path integral fails to converge \cite{GIBBONS1978141}. This reflects the negative-mode and contour subtleties of the dS$_2$ branch.

In \cite{Blacker:2025zca,Prestidge:1999uq}, a partial cure is provided by analytically continuing the radial coordinate and the horizon radius into the complex plane
\begin{equation}\label{eq:complexify}
r_{i}\;\to\; i\rho_{i}\,,\qquad T_{i}\;\to\; iT_{i}\,,
\end{equation}
so that the near-horizon geometry becomes $-(\mathrm{EAdS}_{2}\times S^{2})$.
The overall minus sign now factors out; the relative sign between the two-dimensional part and the sphere is eliminated. 
However, the problem reappears in the neutral limit.
If one adopts the prescription of \cite{Kolanowski:2024zrq,Marolf:2022ntb,Liu:2023jvm} to redefine the inner product, a convergent path integral can be recovered in the neutral limit.
For general cases, however, the simultaneous presence of negative-norm gauge modes and negative effective eigenvalues means the complexification scheme does not furnish a fully convergent Euclidean path integral.
When the near-horizon region contains Mink$_2$, as in the ultracold limit of the RN-dS$_4$ case, the modes take the form \eqref{flat}. Their normalizability and measure require a separate treatment, so the flat branch should not be interpreted as an ordinary zero-eigenvalue limit of the AdS$_2$ determinant\cite{PhysRevD.25.330,Afshar:2021qvi}.

Note that the integrand shown in \eqref{integrand} exhibits a highly nontrivial dependence on the metric function and its derivatives. It involves both $f''(r_i)$ and $f'''(r_i)$, multiplied by intricate rational functions of $y$ and polynomial combinations of the mode number $n$.
One might therefore expect the final eigenvalue shift $\delta \lambda_n$ to carry a complicated imprint of the specific black hole geometry.
Remarkably, however, upon performing the $y$-integration from $1$ to $\infty$, all such dependence on $f''(r_i)$, $f'''(r_i)$ and the coordinate $y$ undergoes a complete cancellation.
Especially, the terms proportional to $f'''(r_i)$, which encode the leading departure from the extremal throat and hence the gluing of the throat into the parent geometry, combine into contributions that do not survive the normalized projection.
The radial integral can be reorganized into an infrared boundary contribution from the asymptotic region of the two-dimensional throat, as made explicit in Section~\ref{sec4}.
What survives is the mode label $|n|$, the small temperature $T$, and the sign determined by the two-dimensional factor. 
Moreover, the final contribution to the partition function is always $\frac{3}{2}\log T$.
This exact cancellation indicates that the eigenvalue shift is governed by the infrared soft sector of the underlying $\mM_2\times \mS^2$ throat rather than by the ultraviolet details of the parent black hole.

\subsection{More examples}

The results established above furnish a general prescription for evaluating zero-mode eigenvalue shifts and the associated partition-function contributions in diverse parent black hole geometries.
In this subsection, we apply this formalism to several black hole solutions in their respective extremal limits.
Through these case-by-case studies, we demonstrate that the mechanism identified previously is universally applicable.
The eigenvalue shifts are dictated by the symmetry structure of the throat geometry, independent of the microscopic specifics of the underlying spacetime.

\subsubsection{Spherically symmetric black holes}

The most straightforward cases are the static spherically symmetric ones represented directly by metric \eqref{s-metric}, where the metric function $f(r)$ characterizes the geometry. We briefly discuss the Schwarzschild and RN black holes in asymptotically flat, AdS, or dS spacetime.

For the neutral case, the black hole has an extremal limit only when the asymptotic spacetime is dS. For the Schwarzschild-dS black hole, only the outer and cosmological horizons exist, so the Nariai limit applies. The near-horizon geometry is dS$_2\times \mS^2$, and the corresponding analytically continued geometry is $(-\mathrm{EAdS}_{2})\times \mS^{2}$. Repeating the analysis outlined above yields the eigenvalue shift from the zero modes of the tensor fluctuation. To leading order in the small temperature $T$, the eigenvalue shift is
\be
\delta \lambda_n =-\frac{|n|T}{16G\cdot r_N},
\ee
where $r_N$ is the horizon radius in the Nariai limit. Thus, for the Nariai limit of the Schwarzschild-dS black hole, we have $\varepsilon = -1$.

For the charged case, we have demonstrated the RN-dS case along with the general mechanism. For the RN black hole in asymptotically flat or AdS spacetime, there are two horizons, the inner and outer black hole horizons. In the cold limit, the near-horizon geometry can be written as a direct product of AdS$_2\times\mS^2$. Although the AdS radius and the geometry when turning on a small temperature are changed, the final eigenvalue shift remains the same. Carrying out the analogous calculation, we obtain
\begin{equation}
    \delta \lambda_n =\frac{|n|T}{16G\cdot r_C},
\end{equation}
as the leading-order correction in temperature $T$. Hence, for the RN case in the cold limit with horizon radius $r_c$, the value $\varepsilon$ characterizing the eigenvalue shift is $\varepsilon =1$.

As discussed before, there can also be a Nariai limit for the RN-dS case, where the black hole horizon and the cosmological horizon coincide. In this limit, the eigenvalue shift is 
\begin{equation}
    \delta \lambda_n =-\frac{|n|T}{16G\cdot r_N},
\end{equation}
so $\varepsilon=-1$, and the other details of the eigenvalue shift are the same.

As a summary, the eigenvalue shift with a small temperature for all of the static spherically symmetric cases can be written as
\begin{equation}
	\delta \lambda_n =\frac{\varepsilon}{16G}\cdot \frac{|n|T}{r_i},
\end{equation}
with the values of $\varepsilon$ shown in Table \ref{tab:SSsign1}.

\begin{table}[H]
    \centering
    \begin{tabular}{c|ccccc}
    \hline\hline
    {Extremal limits} & \makecell{S-dS \\ Nariai limit} & \makecell{RN \\ Cold limit} &\makecell{RN-AdS \\ Cold limit} & \makecell{RN-dS \\ Nariai limit} & Ultracold limit\\ \hline 
    $\varepsilon$ & $-$ & $+$ & $+$ & $-$ & 0  \\ \hline\hline
    \end{tabular}
    \caption{The value of $\varepsilon$ for various spherically symmetric black holes.}
    \label{tab:SSsign1}
\end{table}

\subsubsection{Rotating black holes}

For rotating configurations, the axisymmetry and frame-dragging effects introduce additional complexity, yet the underlying decoupling mechanism in the extreme limits remains parallel to the spherically symmetric cases.

In this subsection, we do not analyze the full Lichnerowicz spectrum of a generic rotating black hole.
We restrict to the near-horizon reparametrization tensor sector whose nonzero components lie along the two-dimensional throat.
Within this sector, the angular warp factor modifies the normalization and projection of the modes, but does not change the radial soft-mode equation.
The universal $|n|T$ dependence therefore persists, while rotation only changes the overall angular factor.
So, the relative factor between the rotating and spherically symmetric results can be traced to the angular part of the normalized matrix element.
In the rotating near-horizon geometry, the two-dimensional metric is multiplied by an angular warp factor denoted by $\sigma$ in \eqref{sigma}.

The normalization factor $N_n$ shown in \eqref{hmn0} comes from the orthonormality of $h_{\mu\nu}$ shown in \eqref{deltamn}
\begin{equation}
	\langle h^{(n)}|h^{(m)}\rangle = \delta^{nm}\,.
\end{equation}
The four-dimensional volume element contains an extra angular warp factor; thus, the normalization factor also has extra dependence on $\sigma$. 
The normalization factor $N_n$ is modified through the angular projection as
\begin{equation}
N_n^2 ~\to~  N_n^2\left[\int_0^\pi \sigma(\theta)\sin\theta\,\dd\theta\right]
\end{equation}
due to the $\sigma$ dependence in the volume element.
On the other hand, the eigenvalue shift defined as an integral might also have angular dependence. The eigenvalue shift defined in \eqref{Ecorrection} can be denoted as
\begin{equation}
    \delta\lambda_n=\langle h^{(n)}|\delta\mathcal D_L|h^{(n)}\rangle.
\end{equation}
The two-dimensional Lichnerowicz deformation scales as $(\delta\mathcal{D}_L)_{\mathrm{rot}}=\sigma^{-1}(\delta\mathcal{D}_L)_{\mathrm{sat}}$, because $\sigma(\theta)$ is constant along the AdS$_2$ directions.
This extra $\sigma^{-1}$ cancels the $\sigma$ left by the angular part of the volume element in the integral.
Consequently, the angular factor in \eqref{Ecorrection} is the unweighted one
\begin{equation}
    \int_0^\pi \,\sin\theta ~\dd\theta,
\end{equation}
whereas the angular factor in the normalization $N_n^2$ has extra
\begin{equation}
    \int_0^\pi \sigma(\theta)\,\sin\theta\,\dd\theta.
\end{equation}
Therefore, the normalized angular projection factor is
\begin{equation}
    \frac{\int_0^\pi \,\sin\theta ~\dd\theta}{\int_0^\pi \sigma(\theta)\,\sin\theta\,\dd\theta}.
\end{equation}
Thus, when a small temperature is turned on, the eigenvalue shift of the Kerr-type rotating examples in this tensor sector can be written as
\begin{equation}
	\delta \lambda_n=\frac{1}{\langle\sigma\rangle} \cdot \frac{\varepsilon}{16G}\cdot\frac{|n|T}{r_i}\,.
\end{equation}
with the angular projection factor defined as
\begin{equation}
	\langle\sigma\rangle=    \frac{\int_0^\pi \,\sigma(\theta)\sin\theta ~\dd\theta\,}
         {\int_0^\pi \,\sin\theta~\dd\theta}\,.\label{sigma-define}
\end{equation}
As discussed in Section \ref{bkg}, the angular warp factor can be expressed as
\begin{equation}
	\sigma(\theta)=1+\frac{a^2}{r_i^2}\cos^2\theta\,.
\end{equation}
For the Kerr-type warp factor, the angular projection factor is explicitly
\begin{equation}
	\frac{1}{\langle\sigma\rangle}=    \frac{\int_0^\pi \,\sin\theta~\dd\theta}{\int_0^\pi \,(1+a^2 \cos^2\theta/r_i^2)\sin\theta ~\dd\theta\,}
         =\frac{3r_i^2}{3r_i^2+a^2}\,.
\end{equation}
This factor is an angular projection factor rather than a modification of the radial soft-mode dynamics.
Thus, the eigenvalue shift for the Kerr-type cases can be written as
\begin{equation}
    \delta\lambda_n
    =\frac{1}{\langle\sigma\rangle} \cdot \frac{\varepsilon}{16G}\cdot\frac{|n|T}{r_i} = \frac{\varepsilon}{16G} \cdot \frac{3r_i \cdot |n| T}{(a^2+3r_i^2)}, \qquad |n|\ge 2.\label{rotating}
\end{equation}

For the asymptotically flat Kerr spacetime, there is only a cold limit, and we can use $r_C$ to denote the event horizon radius in the cold limit.
For the Kerr case, we also have $a=r_C$ in the extremal limit, so we have
\begin{equation}
	\sigma(\theta)=1+\cos^2\theta\,.
\end{equation}
The angular projection factor can be calculated as
\begin{equation}
	\frac{1}{\langle\sigma\rangle}=    \frac{\int_0^\pi \,\sin\theta~\dd\theta}{\int_0^\pi \,(1+\cos^2\theta)\sin\theta ~\dd\theta\,}
         =\frac{3}{4}\,,
\end{equation}
and the eigenvalue shift for the Kerr case can be written as
\begin{equation}
    \delta\lambda_n 
    =\frac{1}{\langle\sigma\rangle} \cdot \frac{1}{16G}\cdot\frac{|n|T}{r_C} = \frac{3}{64G}\cdot \frac{|n|T}{r_C}, \qquad |n|\ge 2.\label{Kerr}
\end{equation}
The result can be regarded as the $a=r_C$ limit of \eqref{rotating} and matches the one derived in \cite{Kapec:2023ruw}. 
As can be seen from \eqref{Kerr}, we have $\varepsilon =+1$ in the eigenvalue shift in the cold limit of the Kerr black hole, and the spectrum is merely changed by a factor when comparing with the spherical cases.
This shows that the factor $3/4$ is not a change in the AdS$_2$ soft spectrum itself, but arises from the way the universal two-dimensional tensor zero modes are embedded and normalized in the warped and fibered four-dimensional Kerr throat.

The Kerr black hole with AdS or dS asymptotics can be calculated similarly. 
We can consider the Kerr-AdS and Kerr-dS cases together.
In the canonical ensemble, the leading correction to the mass is quadratic in $T$, and the expansion follows a similar pattern. 
For Kerr-(A)dS, we have
\begin{equation}
    \Delta_r = (r^2+a^2)\left(1\pm \frac{r^2}{\ell^2} \right)-2GMr ,\label{Delta-r}
\end{equation}
where we have a plus sign for the asymptotically AdS spacetime and a minus sign for the dS case in \eqref{Delta-r}.
The extremality conditions $\Delta_r(r_i)=0$, and $\Delta_r'(r_i)=0$ give
\begin{equation}
    a^2= r_i^2 \frac{\ell^2\pm 3r_i^2}{\ell^2\mp r_i^2}.
\end{equation}
Substituting this relation into \eqref{sigma-define}, we have
\begin{equation}
    \frac{1}{\langle\sigma\rangle}= \frac{3}{4} \left(1\mp\frac{r_i^2}{\ell^2}\right).
\end{equation}
So, the eigenvalue shift for the Kerr-(A)dS black in the near-extremal limit can be obtained as
\be
\delta \lambda_n = \frac{3\varepsilon}{64G} \left(1-e\cdot \frac{r_i^2}{\ell^2}\right)\cdot \frac{|n|T}{r_i}, \quad\quad |n| \geq 2.
\ee
Note that we can write the Kerr black hole with different asymptotics together with the introduced $e$.
Here, $e$ can take the values $+1$, $-1$, and $0$, corresponding to the AdS, dS, and flat asymptotics.
$r_i$ denotes the degenerate horizon radius of the corresponding extremal branch.
The sign of the eigenvalue shift is still determined by whether the two-dimensional throat is locally AdS$_2$ or dS$_2$; the factor $\langle\sigma\rangle^{-1}$ only encodes the angular normalization of the rotating throat.

For the cold limit of Kerr black holes, no matter what the asymptotic geometry is, we always have 
\be
\delta \lambda_n = \frac{3}{64G} \left(1-e\cdot \frac{r_C^2}{\ell^2}\right)\cdot \frac{|n|T}{ r_C}, \quad\quad |n| \geq 2.
\ee
with $\varepsilon = +1$. 
In the Nariai limit of the Kerr-dS black hole, the sign of the eigenvalue shift is different; one has
\be
\delta \lambda_n =- \frac{3}{64G}\left(1+ \frac{r_N^2}{\ell^2}\right)\cdot \frac{|n|T}{ r_N}, \quad\quad |n| \geq 2,
\ee
with horizon radius in the Nariai limit denoted as $r_N$. So, for the Nariai limit of the Kerr-dS spacetime, the sign of the eigenvalue shift is $\varepsilon = -1$.
The results for the Kerr-(A)dS cases can be compared with the results derived in \cite{Maulik:2024dwq,Maulik:2025mmt,Blacker:2025zca}.

For the Kerr-Newman black hole with different asymptotics, one can derive the relation between $a$ and $r_i$ as in the previous cases, substitute it into the angular warp factor, and work out the angular projection factor $\langle\sigma\rangle^{-1}$. Note that there are only two equations $\Delta_r=\Delta_r'=0$ determining the relation between $a$ and $r_i$; however, there are more than two parameters for the Kerr-Newman black holes. Therefore, this relation cannot be completely fixed without specifying additional parameters.
So one can keep 
\begin{equation}
    \delta\lambda_n= \frac{\varepsilon}{16G} \cdot \frac{3r_i \cdot |n| T}{(a^2+3r_i^2)}, \qquad |n|\ge 2,\label{KN1}
\end{equation}
as the result of the Kerr-Newman black hole with different asymptotics. It is clear that the RN result can be recovered as $a\to 0$.
Or one can use the electric charge $q$ to express the eigenvalue shift. It can be shown that in the extremal limit, we have
\begin{equation}
	a^2=\frac{r_i^2(\ell^2\pm 3r_i^2)- q^2\ell^2}{\ell^2\mp  r_i^2}\,.
\end{equation}
So the eigenvalue shift can be written as
\begin{equation}
    \delta\lambda_n= \frac{\varepsilon}{16G} \cdot \frac{3}{4-q^2/r_i^2} \left(1-e\cdot \frac{r_i^2}{\ell^2}\right)\cdot \frac{|n|T}{ r_i}, \qquad |n|\ge 2.\label{KN2}
\end{equation}
The Kerr results can be recovered in the $q\to 0$ limit.
Note that it is obviously \eqref{KN1} and \eqref{KN2} are equivalent, and The cases with zero spinning parameter and zero charge can all be recovered from \eqref{KN1}. The result for the Kerr-Newman case is consistent with the one derived in \cite{Maulik:2024dwq}.

So for the stationary rotating black hole case, the eigenvalue shift from zero modes in the near-extremal limit can be written as
\be
\delta \lambda_n = \frac{\varepsilon}{16G} \cdot \frac{3r_i \cdot |n| T}{(a^2+3r_i^2)}, \qquad |n| \geq 2.
\ee
The Kerr and RN cases can be recovered within the appropriate limits. The parameter $\varepsilon$ characterizes the sign of the eigenvalue shift. 
In the examples considered here, $\varepsilon = +1$ in the cold limit, $\varepsilon = -1$ in the Nariai limit, and $\varepsilon = 0$ in the ultracold limit.
A more explicit demonstration of $\varepsilon$ of the rotating cases is presented in Table~\ref{tab:Rsign}.

\begin{table}[H]
    \centering
    \begin{tabular}{c|ccccc}
    \hline\hline
    {Extremal limits} & \makecell{Kerr (dS or AdS) \\ Cold limit} & \makecell{Kerr-dS \\ Nariai limit} & \makecell{Kerr-Newman-dS \\ Nariai limit}  & Ultracold limit \\ \hline
    $\varepsilon$ & $+$ & $-$ & $-$ & 0 \\\hline\hline
    \end{tabular}
    \caption{The value of $\varepsilon$ for stationary rotating black holes.}
    \label{tab:Rsign}
\end{table}

There are also more situations with additional geometric structures, such as accelerating black holes~\cite{Xu:2025dku} and higher-dimensional Myers-Perry black holes~\cite{Maulik:2024dwq}. 
In these cases, the detailed near-horizon geometry may contain extra warping, fibration, acceleration parameters, or multiple rotation parameters. 
These features can change the overall numerical prefactor in the eigenvalue shift.
Nevertheless, the leading correction always retains the universal form
\begin{equation}
    \delta\lambda_n \propto  \varepsilon\cdot |n|T,
    \qquad |n|\ge2 ,
\end{equation}
The sign $\varepsilon$ is fixed with the two-dimensional near-horizon factor
\begin{equation}
    \varepsilon=+1
    \quad \text{for an } \text{AdS}_2 \text{ throat};
    \qquad
    \varepsilon=-1
    \quad \text{for a } \text{dS}_2 \text{ throat}.
\end{equation}
The parameter $\varepsilon$ encodes the stability of the throat perturbation.
The prefactor is sensitive to the details of the parent geometry, whereas the $|n|T$ scaling follows from the near-horizon soft sector and the lifting of the corresponding zero modes at small temperature. 

%\textcolor{red}{Too early: In this sense, the linear $|n|\cdot T$ behavior is a universal infrared property of the near-horizon soft sector, while the prefactor is ultraviolet data associated with the embedding of the throat into the parent black hole geometry.}

\section{Universality of the near-horizon soft modes}
 \label{sec4}

The simplification demonstrated in previous sections strongly suggests that the eigenvalue shift is controlled by a universal symmetry principle rather than by the microscopic structure of the parent geometry.
In this section, we first summarize this universal result in a compact form, and then explain its physical origin in terms of the infrared Schwarzian soft sector of the near-horizon throat.

\subsection{A proposition for the near-horizon soft modes}

The preceding calculation can be summarized as a conditional statement about the soft tensor sector.
Although the local matrix element of the Lichnerowicz deformation depends on the detailed embedding of the near-horizon throat into the parent black hole geometry, the projected eigenvalue shift is controlled only by the universal soft sector of the near-horizon throat.
The following proposition is summarized with five assumptions; thus, it is not a claim about the full Lichnerowicz spectrum of every black hole background, but about the finite-temperature lifting of a particular near-horizon transverse-traceless sector.

\begin{proposition}[Leading lifting of near-horizon tensor zero modes]
Consider a near-extremal black hole whose extremal near-horizon geometry contains a two-dimensional maximally symmetric factor $\mM_2$, and the remaining angular dependence enters only through a smooth projection factor.
The relevant perturbations being considered are normalizable transverse-traceless tensor modes supported on $\mM_2$, which do not mix with the matter fluctuations at leading order of~$T$.
Then the leading finite-temperature eigenvalue shift takes the form
\begin{equation}
\delta\lambda_n = \frac{1}{\langle\sigma\rangle}\cdot \frac{\varepsilon}{16G}\cdot  \frac{|n|T}{r_i}, \qquad |n|\geq 2\,,
\label{eq:theorem}
\end{equation}
Here, $r_i$ is the degenerate horizon radius and $\langle\sigma\rangle^{-1}$ is the angular projection factor.
For static spherically symmetric cases $\langle\sigma\rangle^{-1}=1$, while in rotating cases $\langle\sigma\rangle^{-1}$ is shown in \eqref{sigma-define}.
The sign parameter is $\varepsilon=+1$ for an AdS$_2$ throat, $\varepsilon=-1$ for a dS$_2$ throat, and $\varepsilon=0$ for the degenerate Mink$_2$ limiting case in which the corresponding normalizable tensor soft sector is absent.
\end{proposition}

The explicit examples discussed above provide nontrivial checks of this proposition.
Spherically symmetric black holes, Kerr-type rotating backgrounds, and the Kerr-Newman family all exhibit the same normalized $|n|\cdot T$ lifting, confirming that the detailed parent-geometry data drop out once the modes are projected and normalized.
Note that, for the static ansatz, this statement follows from the normalized projection~\eqref{lambda-n} and radial integral~\eqref{hDh}. 

The compact parameter $\varepsilon$ should not be understood as implying that the three cases have identical Euclidean path integral status.
The AdS$_2$ case gives the controlled determinant contribution.
The dS$_2$ case involves negative eigenvalues and requires a contour prescription or analytic continuation.
The Mink$_2$ case is a degenerate limit in which the normalizability of the corresponding modes must be treated separately.
Within these assumptions, the parent black hole determines the existence of the throat, the horizon scale $r_i$, the sign of the two-dimensional factor, and possible angular projection factors.
The continuous gluing data, such as near-horizon Taylor coefficients of the parent radial function, enter the unintegrated Lichnerowicz matrix element but do not survive the normalized projection.

\subsection{The bulk-boundary matching with the Schwarzian mode}

The proposition above naturally calls for a symmetry-based interpretation.
From the viewpoint of ordinary perturbation theory, the cancellation of the parent-geometry data appears highly nontrivial, since the unintegrated Lichnerowicz matrix element knows about the detailed gluing of the throat to the ultraviolet region.
However, it is not hard to notice that the modes being lifted are not generic metric fluctuations, but soft tensor modes generated by large diffeomorphisms of the near-horizon geometry, and hence they should probe the infrared reparametrization sector of the throat.
In this subsection, we will show that, in the extremal limit, this sector is associated with an emergent boundary reparametrization symmetry, while a small temperature provides an explicit breaking of that symmetry.
It is therefore natural to interpret the lifted tensor zero modes as pseudo-Goldstone modes, whose small eigenvalue shift is controlled by the Schwarzian soft dynamics rather than by the ultraviolet details of the parent black hole.

The eigenvalue shift $\delta\lambda_n$ is the matrix element of the direct Gaussian kernel $\mathcal D_L$, induced for the lifted zero modes. For a unit-normalized complex mode, its contribution to the quadratic action is $\delta\lambda_n$. 
Accordingly, the effective action in this sector is
\begin{equation}
I_{\eff} =\frac{1}{32\pi G} \int \dd^4 x \sqrt{\bar g}\,h^*_{\alpha\beta}\delta\Delta_L^{\alpha\beta\mu\nu}h_{\mu\nu}.
\end{equation}
For the extremal throat metric shown in \eqref{barg}, one has
\begin{equation}
\sqrt{\bar g} =\frac{2r_i^2}{f''(r_i)}\sin\theta.
\end{equation}
The projected radial part of the effective action can be written as a total derivative, so the contribution is fixed by boundary data of the two-dimensional throat.
After performing the integration along the $(\tau,\theta,\phi)$ directions, one obtains
\begin{equation}
I_{\eff} = T\cdot  N_n^2 \int_1^\infty \dd y\,\frac{\partial \mF(y)}{\partial y}=T \cdot N_n^2 \left[ \mF(\infty)-\mF(1)\right]\,,
\end{equation}
with
\begin{equation}
\begin{aligned}
\mF(y)= -\frac{n^2(n^2-1)^2\pi^2 r_i f''}{2G}\Bigg[&
\frac{|n|-2}{|n|-1} \left(\frac{y-1}{y+1}\right)^{|n|-1}
-2 \left(\frac{y-1}{y+1}\right)^{|n|}\\
&+ \frac{|n|+2}{|n|+1} \left(\frac{y-1}{y+1}\right)^{|n|+1} \Bigg].
\end{aligned}
\end{equation}
The effective action has a direct near-horizon geometry interpretation. 
The coordinate used above is related to the standard radial coordinate of the two-dimensional factor by $y=\cosh\rho$. 
Therefore, $y=1$ corresponds to the center of the Euclidean disk, while $y\rightarrow \infty$ is the asymptotic boundary of the two-dimensional throat.
We have
\begin{equation}
\mF(1)=0, \qquad \mF(\infty) = \frac{\pi^2 r_i}{G} n^2(n^2-1)f''(r_i).
\end{equation}
Therefore, the corresponding quadratic action is
\begin{equation}
I_{\eff} = \frac{\pi^2 r_i N_n^2 T f''(r_i)}{G} n^2(n^2-1).\label{Ieff}
\end{equation}
The nonzero value of $\mathcal F(\infty)$ should therefore be interpreted as an infrared boundary contribution from the asymptotic two-dimensional $\mM_2$.
Using the normalization in~\eqref{Nn},
one obtains $\delta\lambda_n \propto \varepsilon\cdot|n|T$, with $|n|\ge2$.

This structure is precisely what one expects from the infrared dynamics of a nearly AdS$_2$ throat and, with the caveats discussed above, its analytically continued dS$_2$ counterpart.
The asymptotic time reparametrization symmetry is spontaneously broken by the AdS$_2$ background to $SL(2,\mathbb R)$, and is explicitly broken by the small departure from extremality.
The corresponding pseudo-Goldstone mode is governed by the Schwarzian effective action \cite{Maldacena:2016upp}.
The connection with the Schwarzian theory can be made more explicit by looking at the quadratic expansion of the Schwarzian action.

In nearly AdS$_2$ gravity, integrating out the dilaton fixes the two-dimensional geometry to have constant curvature.
The remaining infrared gravitational degree of freedom is not a propagating bulk graviton, but the reparametrization of the asymptotic boundary curve Diff$(S^1)$.

The on-shell gravitational action then reduces to the boundary Schwarzian action \cite{Mertens:2022irh}
\begin{equation}
I_{\rm Sch} \sim C \int \dd\tau\, \text{Sch}[F(\tau),\tau],
\end{equation}
where $C$ is the Schwarzian coupling. 
The Schwarzian derivative is defined by
\begin{equation}
\text{Sch}[F(\tau),\tau] = \frac{F'''(\tau)}{F'(\tau)}-\frac{3}{2} \left(\frac{F''(\tau)}{F'(\tau)}\right)^2 .
\end{equation}
It is invariant under the global $SL(2,\mathbb R)$ transformation, which reflects the fact that the physical Schwarzian modes live in the coset
\begin{equation}
{\mathrm{Diff}(S^1)}/{SL(2,\mathbb R)} .
\end{equation}
Thus, the Schwarzian action describes the pseudo-Goldstone mode associated with the breaking of the near-AdS$_2$ reparametrization symmetry down to $SL(2,\mathbb R)$.
Expanding around the thermal saddle gives, in the boundary reparametrization basis, the quadratic action for fluctuation field $\epsilon$ takes the schematic form
\begin{equation}
I_{\rm Sch}^{(2)} \propto \int d\tau\, \left[ \left(\epsilon''(\tau)\right)^2 - \left(\epsilon'(\tau)\right)^2 \right].
\end{equation}
So, for a Fourier mode $\epsilon(\tau) =  \epsilon_n e^{-in\tau}$, one obtains the quadratic Schwarzian action
\begin{equation}
I_{\rm Sch}^{(2)} \propto T\cdot n^2(n^2-1)\,\bar\epsilon_n\epsilon_n, \qquad |n|\ge2.
\end{equation}
The modes $n=0,\pm1$ are excluded because they generate the unbroken $SL(2,\mathbb R)$ isometries.
Hence, the physical Schwarzian soft modes live in $\mathrm{Diff}(S^1)/SL(2,\mathbb R)$ and are labelled by $|n|\ge2$.

The Schwarzian quadratic kernel in the
$\epsilon_n$-basis is
\begin{equation}
I_{\rm Sch}^{(2)}\propto T\cdot n^2(n^2-1).
\end{equation}
This is precisely the same polynomial dependence on $n$ that appears in the effective action for the near-horizon tensor zero modes obtained in \eqref{Ieff}, before substituting the bulk normalization factor. 
The difference between the Schwarzian expression and the final lifted eigenvalue is therefore not a difference in the underlying soft sector, but comes from the normalization map between the boundary reparametrization variable and the bulk tensor variable.
The Schwarzian action is naturally written in terms of the boundary reparametrization amplitude $\epsilon_n$, whereas the bulk Lichnerowicz problem is written in terms of the normalized metric perturbation
\begin{equation}
h_{ab}^{(n)} = \mathcal L_{\zeta_n}\bar g_{ab}.
\end{equation}
The large diffeomorphism $\zeta_n$ is generated by the same boundary Schwarzian mode $\epsilon_n$.
In this sense, $h^{(n)}_{ab}$ can be regarded as the metric fluctuation generated by $\epsilon_n$, and the difference between the bulk zero modes and the Schwarzian theory is that the bulk zero modes $h^{(n)}_{ab}$ are normalized with $N_n$.

After this normalization is imposed, the $n^2(n^2-1)$ dependence is converted into a linear $|n|$ dependence.
Note that the $T$ dependence is the same because the linear dependence on $T$ follows from ordinary first-order perturbation theory around extremality.
Therefore, the Schwarzian quadratic kernel and the bulk lifting result are two representations of the same near-horizon soft dynamics. 
The former is written in the boundary $\epsilon_n$-basis, while the latter is written in the normalized bulk tensor-mode basis.

This boundary representation explains why the final answer is insensitive to the ultraviolet details of the parent geometry.
Although the unintegrated Lichnerowicz matrix element depends on the ultraviolet gluing data of the parent black hole geometry, all such details cancel after projection onto the soft tensor sector.
The remaining contribution is controlled only by the infrared reparametrization dynamics of the throat, described by the Schwarzian soft mode living in $\mathrm{Diff}(S^1)/SL(2,\mathbb R)$.
The exclusion of the $n=0,\pm1$ modes reflects the unbroken $SL(2,\mathbb R)$ isometries, while the physical modes $|n|\ge2$ acquire a universal lifting proportional to $\varepsilon |n|T$.
Thus, the final result depends only on the horizon scale $r_i$, the mode number $n$, the explicit symmetry-breaking scale $T$, and the sign determined by the two-dimensional throat geometry, within the assumptions of the near-horizon soft tensor sector.
The matching established here is a spectral matching between the quadratic Schwarzian kernel and the normalized Lichnerowicz lifting.
A derivation from the full gravitational symplectic form or from the complete gauge-fixed path integral measure is beyond the scope of the present work.

%%%%%%%%%%%%%%%%%%%%%%%%%%%%%%%%%%%%%%%%%%%%%%%%%%%%%%%%%%%%%%%%%%%%%%%%%%%%%%%%%%%%%%%%%%%%%%%%%%%%

\section{Conclusion and discussion}
\label{con}
%%%%%%%%%%%%%%%%%%%%%%%%%%%%%%%%%%%%%%%%%%%%%%%%%%%%%%%%%%%%%%%%%%%%%%%%%%%%%%%%%%%%%%%%%%%%%%%%%%%%

In this paper, we studied the finite-temperature lifting of near-horizon gravitational tensor zero modes from the Lichnerowicz spectrum of near-extremal black holes. 
For the class of backgrounds specified in Section~\ref{sec4}, the leading shift is an infrared property of the throat.
The main result is that, for the static spherically symmetric geometries considered here, the leading eigenvalue shift of the normalized near-horizon transverse-traceless tensor zero modes takes the universal form
\begin{equation}
\delta\lambda_n = \frac{\varepsilon}{16G}\cdot \frac{|n|T}{r_i}, \qquad |n|\geq 2,
\end{equation}
where $r_i$ is the degenerate horizon radius and $\varepsilon$ is determined by the two-dimensional near-horizon factor. For stationary rotating geometries, the same linear $|n|\cdot T$ dependence persists in the near-horizon reparametrization tensor sector, while the angular warping of the two-dimensional throat modifies only the overall projection factor. In the Kerr-type warped throat discussed above, this gives
\begin{equation}
\delta\lambda_n = \frac{1}{\langle \sigma\rangle}\cdot \frac{\varepsilon}{16G} \cdot \frac{|n|T}{r_i},\qquad |n|\geq 2.
\end{equation}
Thus the parent geometry fixes the horizon scale, the sign of the two-dimensional throat, and possible angular normalization factors, but it does not change the universal dependence on the Fourier label and the temperature.

The nontrivial point is that the local matrix element of the deformed Lichnerowicz operator contains detailed parent-geometry data, including the near-horizon coefficients such as $f''(r_i)$ and $f'''(r_i)$.
After projection onto normalized tensor zero modes, these continuous gluing data cancel.
The final answer remembers only the horizon scale, the temperature, possible angular projection factors, and the sign associated with the two-dimensional factor.
In this sense, the leading finite-temperature lifting is controlled by the infrared soft sector rather than by the ultraviolet completion of the throat.

This spectral result is also the bulk realization of the Schwarzian soft sector.
The tensor zero modes are generated by large near-horizon diffeomorphisms, with the $n=0,\pm1$ modes removed as the exact $SL(2,\mathbb R)$ isometries.
The remaining modes correspond to $\mathrm{Diff}(S^1)/SL(2,\mathbb R)$.
In the boundary reparametrization basis, the quadratic Schwarzian action contains the kernel $T n^2(n^2-1)$.
In the bulk Lichnerowicz problem, the same polynomial appears before imposing the normalization of the transverse-traceless tensor modes.
The normalization map between the boundary reparametrization amplitude and the bulk tensor mode converts this kernel into the linear $|n|\cdot T$ eigenvalue shift.
The determinant consequence follows immediately in the stable AdS$_2$ branch.
The lifted tensor modes give
\begin{equation}
  \delta\log Z = -\frac12\sum_{|n|\ge2}\log\delta\lambda_n
  = \frac{3}{2}\log\frac{T}{r_i}+\cdots\,,
\end{equation}
where the dots include temperature-independent contributions.
This is the spectral origin of the universal logarithmic temperature dependence from the tensor soft sector.

The compact sign parameter $\varepsilon$ must be interpreted with care.
The AdS$_2$ case gives the controlled Euclidean determinant contribution.
The dS$_2$ case has a negative analytically continued coefficient and requires a contour prescription or analytic continuation.
The Mink$_2$ case is a degenerate limit in which the corresponding normalizable tensor soft sector is absent and must be treated separately.
Thus, the AdS$_2$, dS$_2$, and Mink$_2$ branches share the same algebraic lifting structure, but they do not have identical Euclidean path-integral status.

It is useful to clarify the relation between the present proposition and the recent examples, like in \cite{Maulik:2024dwq,Maulik:2025mmt,Blacker:2025zca,Xu:2025dku,Alvarado:2026kio,Acito:2026mmf}, and the universality theorem for the quantum thermodynamics of near-extremal black holes in \cite{PandoZayas:2026vbg}.
In the present work, we address why the leading-order behavior forgets the parent geometry details with an explicit demonstration of the metric function cancellation and a spectral origin explanation.
Our proposition goes beyond those examples by fixing the normalized Lichnerowicz lifting coefficient and identifying the surviving geometric data.
Moreover, the central contribution is to establish the universality of the normalized near-horizon Lichnerowicz spectrum itself and explain its bulk origin.
The local Lichnerowicz matrix element may contain detailed information about the parent geometry, angular warping, and gluing data, but these dependences cancel after projection onto the normalized soft tensor sector.
Thus, the present result turns the examples and the thermodynamic theorem into a sharper Lichnerowicz spectral statement and identifies the lifted tensor spectrum as the bulk realization of the Schwarzian reparametrization soft mode.
%The near-horizon Taylor coefficients carrying local parent-geometry and gluing information cancel from the static matrix element after radial integration and bulk normalization, leaving only the horizon scale.
%In the Kerr-type examples admitting radial-angular factorization, the remaining angular dependence modifies only the overall projection factor.
%We further show that the universal $T|n|$ lifting is the normalized bulk representation of the Schwarzian kernel $Tn^2(n^2-1)$.
The thermodynamic universality is thus traced to a universal soft spectrum governed by the infrared reparametrization dynamics of the throat.

Several comments are in order. 
First, our analysis isolates the tensor zero-mode sector.
A complete one-loop calculation can also involve vector modes, gauge modes, matter modes, ghosts, and possible boundary degrees of freedom.
These additional sectors may change the full one-loop answer, but they do not alter the universal tensor-mode lifting derived here.
Second, for rotating black holes, the universal two-dimensional soft sector is dressed by angular factors.
The examples considered in this work suggest that such angular dependence only modifies the overall prefactor, but a fully general treatment of warped and fibered near-horizon geometries would be useful.
Finally, the cases with $\varepsilon=-1$ and $\varepsilon=0$ require special care in the Euclidean path integral.
The dS$_2$ branch leads to negative lifted eigenvalues and hence to contour subtleties, while the flat branch lacks the same normalizable tensor soft sector.
These questions are natural extensions of the near-horizon universality statement established here.

To summarize, the leading finite-temperature lifting of near-horizon tensor zero modes is universal for a broad class of extremal black holes under the assumptions stated in Section \ref{sec4}. The dependence on $|n|$ and $T$ is fixed by the infrared reparametrization dynamics of the throat. The universal lifting itself is the bulk Lichnerowicz spectral manifestation of the Schwarzian infrared dynamics.

\section*{Acknowledgements}
We thank Jan de Boer for the helpful discussions. 
This work is supported by the National Natural Science Foundation of China (Grant No. 12405073) and the Natural Science Foundation of Tianjin (Grant No. 25JCQNJC01920). PC is partially supported by Tianjin University Self-Innovation Fund Extreme Basic Research Project (Grant No. 2025XJ21-0007).

%%%%%%%%%%%%%%%%%%%%%%%%%%%%%%%%%%%%%%%%%%%%%%%%%%%%%%%%%%%%%%%%%%%%%%%%%%%%%%%%%%%%%%%%%%%%%%%%%%%%

%\bibliography{ref}
%\bibliographystyle{utphys}

\providecommand{\href}[2]{#2}\begingroup\raggedright\endgroup

%%%%%%%%%%%%%%%%%%%%%%%%%%%%%%%%%%%%%%%%%%%%%%%%%%%%%%%%%%%%%%%%%%%%%%%%%%%%%%%%%%%%%%%%%%%%%%%%%%%%

\end{document}